\documentclass[conference,a4paper]{IEEEtran}
\usepackage[dvips]{graphicx,color}
\usepackage{latexsym}
\usepackage{amssymb}
\usepackage{amsmath, bm}
\usepackage{url}
\usepackage{array}

\begin{document}
%
%
%
%
\newcommand{\qed}{\hfill$\square$}
\newcommand{\suchthat}{\mbox{~s.t.~}}
\newcommand{\markov}{\leftrightarrow}
%
%
\newenvironment{pRoof}{%
 \noindent{\em Proof.\ }}{%
 \hspace*{\fill}\qed \\
 \vspace{2ex}}


\newcommand{\ket}[1]{| #1 \rangle}
\newcommand{\bra}[1]{\langle #1 |}
\newcommand{\bol}[1]{\mathbf{#1}}
\newcommand{\rom}[1]{\mathrm{#1}}
\newcommand{\san}[1]{\mathsf{#1}}
\newcommand{\mymid}{:~}
\newcommand{\argmax}{\mathop{\rm argmax}\limits}
\newcommand{\argmin}{\mathop{\rm argmin}\limits}
%
%
%
%
\newcommand{\bc}{\begin{center}}  %
\newcommand{\ec}{\end{center}}
\newcommand{\befi}{\begin{figure}[h]}  %
\newcommand{\enfi}{\end{figure}}
\newcommand{\bsb}{\begin{shadebox}\begin{center}}   %
\newcommand{\esb}{\end{center}\end{shadebox}}
\newcommand{\bs}{\begin{screen}}     %
\newcommand{\es}{\end{screen}}
\newcommand{\bib}{\begin{itembox}}   %
\newcommand{\eib}{\end{itembox}}
\newcommand{\bit}{\begin{itemize}}   %
\newcommand{\eit}{\end{itemize}}
\newcommand{\defeq}{:=}

\newcommand{\Qed}{\hbox{\rule[-2pt]{3pt}{6pt}}}
\newcommand{\beq}{\begin{equation}}
\newcommand{\eeq}{\end{equation}}
\newcommand{\beqa}{\begin{eqnarray}}
\newcommand{\eeqa}{\end{eqnarray}}
\newcommand{\beqno}{\begin{eqnarray*}}
\newcommand{\eeqno}{\end{eqnarray*}}
\newcommand{\ba}{\begin{array}}
\newcommand{\ea}{\end{array}}
\newcommand{\vc}[1]{\mbox{\boldmath $#1$}}
\newcommand{\lvc}[1]{\mbox{\footnotesize \boldmath $#1$}}
\newcommand{\svc}[1]{\mbox{\scriptsize\boldmath $#1$}}
\newcommand{\ssvc}[1]{\mbox{\tiny\boldmath $#1$}}

\newcommand{\wh}{\widehat}
\newcommand{\wt}{\widetilde}
\newcommand{\ts}{\textstyle}
\newcommand{\ds}{\displaystyle}
\newcommand{\scs}{\scriptstyle}
\newcommand{\vep}{\varepsilon}
\newcommand{\rhp}{\rightharpoonup}
\newcommand{\cl}{\circ\!\!\!\!\!-}
\newcommand{\bcs}{\dot{\,}.\dot{\,}}
\newcommand{\eqv}{\Leftrightarrow}
\newcommand{\leqv}{\Longleftrightarrow}
\newtheorem{co}{Corollary} 
\newtheorem{lm}{Lemma} 
\newtheorem{Ex}{Example} 
\newtheorem{Th}{Theorem}
\newtheorem{df}{Definition} 
\newtheorem{pr}{Property} 
\newtheorem{pro}{Proposition} 
\newtheorem{rem}{Remark} 

\newcommand{\lcv}{convex } 

\newcommand{\hugel}{{\arraycolsep 0mm
                    \left\{\ba{l}{\,}\\{\,}\ea\right.\!\!}}
\newcommand{\Hugel}{{\arraycolsep 0mm
                    \left\{\ba{l}{\,}\\{\,}\\{\,}\ea\right.\!\!}}
\newcommand{\HUgel}{{\arraycolsep 0mm
                    \left\{\ba{l}{\,}\\{\,}\\{\,}\vspace{-1mm}
                    \\{\,}\ea\right.\!\!}}
\newcommand{\huger}{{\arraycolsep 0mm
                    \left.\ba{l}{\,}\\{\,}\ea\!\!\right\}}}
\newcommand{\Huger}{{\arraycolsep 0mm
                    \left.\ba{l}{\,}\\{\,}\\{\,}\ea\!\!\right\}}}
\newcommand{\HUger}{{\arraycolsep 0mm
                    \left.\ba{l}{\,}\\{\,}\\{\,}\vspace{-1mm}
                    \\{\,}\ea\!\!\right\}}}

\newcommand{\hugebl}{{\arraycolsep 0mm
                    \left[\ba{l}{\,}\\{\,}\ea\right.\!\!}}
\newcommand{\Hugebl}{{\arraycolsep 0mm
                    \left[\ba{l}{\,}\\{\,}\\{\,}\ea\right.\!\!}}
\newcommand{\HUgebl}{{\arraycolsep 0mm
                    \left[\ba{l}{\,}\\{\,}\\{\,}\vspace{-1mm}
                    \\{\,}\ea\right.\!\!}}
\newcommand{\hugebr}{{\arraycolsep 0mm
                    \left.\ba{l}{\,}\\{\,}\ea\!\!\right]}}
\newcommand{\Hugebr}{{\arraycolsep 0mm
                    \left.\ba{l}{\,}\\{\,}\\{\,}\ea\!\!\right]}}
\newcommand{\HUgebr}{{\arraycolsep 0mm
                    \left.\ba{l}{\,}\\{\,}\\{\,}\vspace{-1mm}
                    \\{\,}\ea\!\!\right]}}

\newcommand{\hugecl}{{\arraycolsep 0mm
                    \left(\ba{l}{\,}\\{\,}\ea\right.\!\!}}
\newcommand{\Hugecl}{{\arraycolsep 0mm
                    \left(\ba{l}{\,}\\{\,}\\{\,}\ea\right.\!\!}}
\newcommand{\hugecr}{{\arraycolsep 0mm
                    \left.\ba{l}{\,}\\{\,}\ea\!\!\right)}}
\newcommand{\Hugecr}{{\arraycolsep 0mm
                    \left.\ba{l}{\,}\\{\,}\\{\,}\ea\!\!\right)}}

\newcommand{\hugepl}{{\arraycolsep 0mm
                    \left|\ba{l}{\,}\\{\,}\ea\right.\!\!}}
\newcommand{\Hugepl}{{\arraycolsep 0mm
                    \left|\ba{l}{\,}\\{\,}\\{\,}\ea\right.\!\!}}
\newcommand{\hugepr}{{\arraycolsep 0mm
                    \left.\ba{l}{\,}\\{\,}\ea\!\!\right|}}
\newcommand{\Hugepr}{{\arraycolsep 0mm
                    \left.\ba{l}{\,}\\{\,}\\{\,}\ea\!\!\right|}}

\newcommand{\MEq}[1]{\stackrel{
{\rm (#1)}}{=}}

\newcommand{\MLeq}[1]{\stackrel{
{\rm (#1)}}{\leq}}

\newcommand{\ML}[1]{\stackrel{
{\rm (#1)}}{<}}

\newcommand{\MGeq}[1]{\stackrel{
{\rm (#1)}}{\geq}}

\newcommand{\MG}[1]{\stackrel{
{\rm (#1)}}{>}}

\newcommand{\MSubeq}[1]{\stackrel{
{\rm (#1)}}{\subseteq}}

\newcommand{\MSupeq}[1]{\stackrel{
{\rm (#1)}}{\supseteq}}

\newcommand{\MPreq}[1]{\stackrel{
{\rm (#1)}}{\preceq}}

\newcommand{\MSueq}[1]{\stackrel{
{\rm (#1)}}{\succeq}}

\newcommand{\Ch}{{\Gamma}}
\newcommand{\Rw}{{W}}

\newcommand{\Cd}{{\cal R}_{\rm d}(\Ch)}
\newcommand{\CdB}{{\cal R}_{\rm d}^{\prime}(\Ch)}
\newcommand{\CdBB}{{\cal R}_{\rm d}^{\prime\prime}(\Ch)}

\newcommand{\Cdi}{{\cal R}_{\rm d}^{\rm (in)}(\Ch)}
\newcommand{\Cdo}{{\cal R}_{\rm d}^{\rm (out)}(\Ch)}

\newcommand{\tCdi}{\tilde{\cal R}_{\rm d}^{\rm (in)}(\Ch)}
\newcommand{\tCdo}{\tilde{\cal R}_{\rm d}^{\rm (out)}(\Ch)}
\newcommand{\hCdo}{  \hat{\cal R}_{\rm d}^{\rm (out)}(\Ch)}

\newcommand{\Cs}{{\cal R}_{\rm s}(\Ch)}
\newcommand{\CsB}{{\cal R}_{\rm s}^{\prime}(\Ch)}
\newcommand{\CsBB}{{\cal R}_{\rm s}^{\prime\prime}(\Ch)}

\newcommand{\Csi}{{\cal R}_{\rm s}^{\rm (in)}(\Ch)}
\newcommand{\Cso}{{\cal R}_{\rm s}^{\rm (out)}(\Ch)}
\newcommand{\tCsi}{\tilde{\cal R}_{\rm s}^{\rm (in)}(\Ch)}
\newcommand{\tCso}{\tilde{\cal R}_{\rm s}^{\rm (out)}(\Ch)}
\newcommand{\cCsi}{\check{\cal R}_{\rm s}^{\rm (in)}(\Ch)}
\newcommand{\Cds}{{\cal C}_{\rm ds}(\Ch)}
\newcommand{\Cdsi}{{\cal C}_{\rm ds}^{\rm (in)}(\Ch)}
\newcommand{\Cdso}{{\cal C}_{\rm ds}^{\rm (out)}(\Ch)}
\newcommand{\tCdsi}{\tilde{\cal C}_{\rm ds}^{\rm (in)}(\Ch)}
\newcommand{\tCdso}{\tilde{\cal C}_{\rm ds}^{\rm (out)}(\Ch)}
\newcommand{\hCdso}{\hat{\cal C}_{\rm ds}^{\rm (out)}(\Ch)}
\newcommand{\Css}{{\cal C}_{\rm ss}(\Ch)}
\newcommand{\Cssi}{{\cal C}_{\rm ss}^{\rm (in)}(\Ch)}
\newcommand{\Csso}{{\cal C}_{\rm ss}^{\rm (out)}(\Ch)}
\newcommand{\tCssi}{\tilde{\cal C}_{\rm ss}^{\rm (in)}(\Ch)}
\newcommand{\tCsso}{\tilde{\cal C}_{\rm ss}^{\rm (out)}(\Ch)}
\newcommand{\Cde}{{\cal R}_{\rm d1e}(\Ch)}
\newcommand{\Cdei}{{\cal R}_{\rm d1e}^{\rm (in)}(\Ch)}
\newcommand{\Cdeo}{{\cal R}_{\rm d1e}^{\rm (out)}(\Ch)}
\newcommand{\tCdei}{\tilde{\cal R}_{\rm d1e}^{\rm (in)}(\Ch)}
\newcommand{\tCdeo}{\tilde{\cal R}_{\rm d1e}^{\rm (out)}(\Ch)}
\newcommand{\hCdeo}{  \hat{\cal R}_{\rm d1e}^{\rm (out)}(\Ch)} 
\newcommand{\Cse}{{\cal R}_{\rm s1e}(\Ch)}
\newcommand{\Csei}{{\cal R}_{\rm s1e}^{\rm (in)}(\Ch)}
\newcommand{\Cseo}{{\cal R}_{\rm s1e}^{\rm (out)}(\Ch)}
\newcommand{\tCsei}{\tilde{\cal R}_{\rm s1e}^{\rm (in)}(\Ch)}
\newcommand{\tCseo}{\tilde{\cal R}_{\rm s1e}^{\rm (out)}(\Ch)}

\newcommand{\Capa}{C}

\newcommand{\ZeTa}{\zeta(S;Y,Z|U)}
\newcommand{\ZeTaI}{\zeta(S_i;Y_i,Z_i|U_i)}

\newcommand{\SP}{\mbox{{\scriptsize sp}}}
\newcommand{\mSP}{\mbox{{\scriptsize sp}}}
\newcommand{\CEreg}{\irBr{rate} }
\newcommand{\CEregB}{rate\MarkOh{-equivocation }}

\newcommand{\Cls}{class NL}
\newcommand{\vSpa}{\vspace{0.3mm}}
\newcommand{\Prmt}{\zeta}
\newcommand{\pj}{\omega_n}

\newfont{\bg}{cmr10 scaled \magstep4}
\newcommand{\bigzerol}{\smash{\hbox{\bg 0}}}
\newcommand{\bigzerou}{\smash{\lower1.7ex\hbox{\bg 0}}}
\newcommand{\nbn}{\frac{1}{n}}
\newcommand{\ra}{\rightarrow}
\newcommand{\la}{\leftarrow}
\newcommand{\ldo}{\ldots}
\newcommand{\ep}{\epsilon }
\newcommand{\typi}{A_{\epsilon }^{n}}
\newcommand{\bx}{\hspace*{\fill}$\Box$}
\newcommand{\pa}{\vert}
\newcommand{\ignore}[1]{}
\newcommand{\nth}{(n)}
\newcommand{\VarXun}{X^n}
\newcommand{\VarYun}{Y^n}
\newcommand{\VarZum}{Y^m}

\newcommand{\VarX}{X}
\newcommand{\VarY}{Y}
\newcommand{\VarZ}{Z}

\newcommand{\calVarX}{{\cal X}}
\newcommand{\calVarY}{{\cal Y}}
\newcommand{\calVarZ}{\cal Z}

\newcommand{\calVarXun}{{\cal X}^n}
\newcommand{\calVarYun}{{\cal Y}^n}
\newcommand{\calVarZum}{{\cal Z}^m}

\newcommand{\varx}{x}
\newcommand{\varxun}{x^n}
\newcommand{\varxa}{x_{1}}
\newcommand{\varxb}{x_{2}}
\newcommand{\varxn}{x_{n}}
\newcommand{\varxt}{x_{t}}

\newcommand{\VarXa}{X_{1}}
\newcommand{\VarXb}{X_{2}}
\newcommand{\VarXn}{X_{n}}
\newcommand{\VarXt}{X_{t}}

\newcommand{\vary}{y}
\newcommand{\varyun}{y^n}
\newcommand{\varya}{y_{1}}
\newcommand{\varyb}{y_{2}}
\newcommand{\varyn}{y_{n}}
\newcommand{\varyt}{y_{t}}

\newcommand{\VarYa}{Y_{1}}
\newcommand{\VarYb}{Y_{2}}
\newcommand{\VarYn}{Y_{n}}
\newcommand{\VarYt}{Y_{t}}

\newcommand{\varz}{z}
\newcommand{\varzum}{z^m}
\newcommand{\varza}{z_1}
\newcommand{\varzb}{z_2}
\newcommand{\varzm}{z_m}
\newcommand{\varzt}{z_t}

\newcommand{\VarZa}{Z_1}
\newcommand{\VarZb}{Z_2}
\newcommand{\VarZm}{Z_m}
\newcommand{\VarZt}{Z_t}

\newcommand{\tX}{\tilde{X}}
\newcommand{\tY}{\tilde{Y}}
\newcommand{\tZ}{\tilde{Z}}
\newcommand{\tZum}{\tilde{Z}^m}

\newcommand{\E}{\mbox{\bf E}}
\newcommand{\Var}{\mbox{\bf Var}}  

\newcommand{\Dist}[1]{{#1}}

\newcommand{\OMeg}[1]{\tilde{Y}_{#1}^m}

\newcommand{\Vcx}{\mbox{\boldmath $x$}}
\newcommand{\Vcy}{\mbox{\boldmath $y$}}



%
%

%
%
%

\newenvironment{jenumerate}
	{\begin{enumerate}\itemsep=-0.25em \parindent=1zw}{\end{enumerate}}
\newenvironment{jdescription}
	{\begin{description}\itemsep=-0.25em \parindent=1zw}{\end{description}}
\newenvironment{jitemize}
	{\begin{itemize}\itemsep=-0.25em \parindent=1zw}{\end{itemize}}
\renewcommand{\labelitemii}{$\cdot$}

\newcommand{\iro}[2]{{\color[named]{#1}#2\normalcolor}}
\newcommand{\irr}{\empty}

\newcommand{\irg}[1]{{\color[named]{Green}#1\normalcolor}}

\newcommand{\irb}{\empty}

\newcommand{\irBl}[1]{{\color[named]{Black}#1\normalcolor}}
\newcommand{\irWh}[1]{{\color[named]{White}#1\normalcolor}}

\newcommand{\irY}[1]{{\color[named]{Yellow}#1\normalcolor}}

\newcommand{\irO}{\empty}
\newcommand{\irBr}{\empty}

\newcommand{\IrBr}[1]{{\color[named]{Purple}#1\normalcolor}}
\newcommand{\irBw}[1]{{\color[named]{Brown}#1\normalcolor}}
\newcommand{\irPk}[1]{{\color[named]{Magenta}#1\normalcolor}}
\newcommand{\irCb}[1]{{\color[named]{CadetBlue}#1\normalcolor}}

\newcommand{\irMho}[1]{{\color[named]{Mahogany}#1\normalcolor}}
\newcommand{\irOlg}[1]{{\color[named]{Black}#1\normalcolor}}

\newcommand{\irBg}[1]{{\color[named]{BlueGreen}#1\normalcolor}}
\newcommand{\irCy}[1]{{\color[named]{Cyan}#1\normalcolor}}
\newcommand{\irRyp }[1]{{\color[named]{RoyalPurple}#1\normalcolor}}

\newcommand{\irAqm}[1]{{\color[named]{Aquamarine}#1\normalcolor}}
\newcommand{\irRyb}[1]{{\color[named]{RoyalBule}#1\normalcolor}}
\newcommand{\irNvb}[1]{{\color[named]{NavyBlue}#1\normalcolor}}
\newcommand{\irSkb}[1]{{\color[named]{SkyBlue}#1\normalcolor}}
\newcommand{\irTeb}[1]{{\color[named]{TeaBlue}#1\normalcolor}}
\newcommand{\irSep}[1]{{\color[named]{Sepia}#1\normalcolor}}
\newcommand{\irReo}[1]{{\color[named]{RedOrange}#1\normalcolor}}
\newcommand{\irRur}[1]{{\color[named]{RubineRed}#1\normalcolor}}
\newcommand{\irSa }[1]{{\color[named]{Salmon}#1\normalcolor}}
\newcommand{\irAp}[1]{{\color[named]{Apricot}#1\normalcolor}}

%
\newenvironment{indention}[1]{\par
\addtolength{\leftskip}{#1}\begingroup}{\endgroup\par}
%
\newcommand{\namelistlabel}[1]{\mbox{#1}\hfill} 
\newenvironment{namelist}[1]{%
\begin{list}{}
{\let\makelabel\namelistlabel
\settowidth{\labelwidth}{#1}
\setlength{\leftmargin}{1.1\labelwidth}}
}{%
\end{list}}
%
%
\newcommand{\bfig}{\begin{figure}[t]}
\newcommand{\efig}{\end{figure}}
\setcounter{page}{1}

\newtheorem{theorem}{Theorem}

\newcommand{\ExP}{2}
\newcommand{\Ep}{\mbox{\rm e}}

\newcommand{\Exp}{\exp
}
\newcommand{\idenc}{{\varphi}_n}
\newcommand{\resenc}{
{\varphi}_n}
\newcommand{\ID}{\mbox{\scriptsize ID}}
\newcommand{\TR}{\mbox{\scriptsize TR}}
\newcommand{\Av}{\mbox{\sf E}}

\newcommand{\Vl}{|}
\newcommand{\Ag}{(R,P_{X^n}|W^n)}
\newcommand{\Agv}[1]{({#1},P_{X^n}|W^n)}
\newcommand{\Avw}[1]{({#1}|W^n)}

\newcommand{\Jd}{X^nY^n}
\newcommand{\IdR}{r_n}

\newcommand{\Index}{{n,i}}

\newcommand{\cid}{C_{\mbox{\scriptsize ID}}}
\newcommand{\cida}{C_{\mbox{{\scriptsize ID,a}}}}
\newcommand{\rmOH}{\empty
}

\newcommand{\NoizeVar}{\sigma^2}

\newcommand{\GN}{
\frac{{\rm e}^{-\frac{(y-x)^2}{2\NoizeVar}}}
{\sqrt{2\pi {\NoizeVar}}}}

\arraycolsep 0.5mm
\date{}
%
\title{
New Converse Bounds for Discrete Memoryless Channels 
in the Finite Blocklength Regime 
}
\author{%
	\IEEEauthorblockN{Yasutada Oohama}
	\IEEEauthorblockA{University of Electro-Communications, 
        Tokyo, Japan\\ 
	Email: \url{oohama@uec.ac.jp}}
%

\thanks{
Y. Oohama is with 
Dept. of Communication Engineering and Informatics,
University of Electro-Communications,
1-5-1 Chofugaoka Chofu-shi, Tokyo 182-8585, Japan.
}%
\thanks{
}
}
\markboth{
}
{
}
\maketitle

\begin{abstract}
We study the determination problem of the channel capacity 
for the discrete memoryless channels in the finite blocklength regime. 
We derive explicit lower and upper bounds of the capacity.
We shall demonstrate that the information spectrum approach 
is quite useful for investigating this problem.
\end{abstract}


\newcommand{\ReF}{

}

\newcommand{\ReFz}{

}

\section{Introduction} 


In this paper we consider the determination problem of the channel 
capacity for the discrete memoryless channels in the finite blocklength
regime. This problem, including the study on the second order coding 
theorems originated from the work by Strassen \cite{Strassen62}, 
has intensively been investigated by 
\cite{Hayashi10}-\cite{Kostina13}.
 
In this paper we propose a new method for the proof of the converse 
coding theorem. Our method is a combination of the information spectrum 
method introduced by Han \cite{Han98InfSpec} and the method of type 
developed by Csisz\'ar and K\"orner \cite{ck}. 

We first generalize a meta converse lemma used in the proof of the 
converse coding theorem in the information spectrum method. Coupling the 
generalized lemma with the method of types, we derive  new converse 
bounds. Those bounds have forms obviously matching achievable bounds 
for sufficiently large code block length. 


\section{
The Capacity of the Discrete Memoryless Channels
}

We consider a discrete memoryless channel(DMC) with the input set 
${\cal X}$ and the output set ${\cal Y}$. We assume that 
${\cal X}$ and ${\cal Y}$ are finite sets. 
The DMC is specified by the following stochastic matrix:
\beq
{W} \defeq \{ {W}(y|x)\}_{
(x,y) 
\in    {\cal X}
\times {\cal Y} 
}.
\eeq
Let $X^n$ be a random variable taking values in ${\cal X}^n$. 
We write an element of ${\cal X}^n$ as 
${\Vcx}=x_{1}x_{2}$$\cdots x_{n}.$ 
Suppose that $X^n$ has a probability distribution on ${\cal X}^n$ 
denoted by 
$p_{X^n}=$ 
$\left\{p_{X^n}({\Vcx} ) 
\right\}_{{{\lvc x} } \in {\cal X}^n}$.
Similar notations are adopted for other random variables. 
Let $Y^n \in {\cal Y}^n$ be a random variable 
obtained as the channel output by connecting 
$X^n$ to the input of channel. We write a conditional 
distribution of $Y^n$ on given $X^n$ as 
$$
W^n=
\left\{W^n({\Vcy} |{\Vcx} )\right\}_{({\lvc x} ,{\lvc y} ) 
\in {\cal X}^n \times {\cal Y}^n}.
$$
Since the channel is memoryless, we have 
\beq
W^n({\Vcy}|{\Vcx} )=\prod_{t=1}^nW (y_t|x_t).
\label{eqn:sde0}
\eeq
Let $K_n$ be uniformly distributed random variables 
taking values in message sets ${\cal K}_n $. 

The random variable $K_n$ is a message sent to the receiver.
A sender transforms $K_n$ into a transmitted sequence 
$X^n$ using an encoder function and sends it to the receiver. 
In this paper we assume that the encoder function $\varphi^{(n)}$ 
is a deterministic encoder. In this case, $\varphi^{(n)}$ is 
is a one-to-one mapping from ${\cal K}_n$ into ${\cal X}^n$. 
\newcommand{\Zapcc}{
a stochastic matrix given by
In this paper we assume that the encoder function $\varphi^{(n)}$ 
we consider two cases on the encoder function. 
One is a case where the encoder function is a deterministic encoder 
denoted by $\varphi^{(n)}$. The other is a case where the encoder 
function is a stochastic encoder denoted by $\varphi_{\rm s}$. 
In the former case 
In the latter case, 
$\varphi_{\rm s}^{(n)}$ is 
a stochastic matrix given by
$$
\varphi_{\rm s}^{(n)}=\{
\varphi_{\rm s}^{(n)}({\Vcx} |k)\}_{
(k,{\Vcx} )\in {\cal K}_n \times {\cal X}^n},
$$ 
where $\varphi_{\rm s}^{(n)}({\Vcx} |k)$ is a conditional probability 
of ${\Vcx}  \in {\cal X}^n$ given message $k\in$
${\cal K}_n$. In the case of deterministic encoders 
In the case of stochastic encoders 
the joint probability mass function on 
$
{\cal X}^n$ 
$\times {\cal Y}^n$ 
is given by
\beqno
& &\Pr\{(K_n,X^n,Y^n)=(k,{\Vcx} ,{\Vcy} )\}
\\
&&= \frac{\varphi_{\rm s}^{(n)}({\Vcx} |k)}{\pa{\cal K}_n\pa }
\prod_{t=1}^n W\left(y_t\left|x_t\right.\right)
\eeqno
where $\pa {\cal K}_n \pa$ is a cardinality 
of the set ${\cal K}_n$. 
}
The joint probability mass function on 
$
{\cal X}^n$ 
$\times {\cal Y}^n$ 
is given by
\beqno
& &\Pr\{(X^n,Y^n)=({\Vcx}, {\Vcy} )\}
=\frac{1}{\pa{\cal K}_n\pa }
\prod_{t=1}^n W\left(y_t\left|x_t(k)\right.\right),
\eeqno
where $x_t(k)=[\varphi^{(n)}(k)]_t$, $t=1,2,\cdots,n$ 
are the $t$-th components of ${\Vcx} ={\Vcx} (k)$ $=\varphi^{(n)}(k)$
and $\pa {\cal K}_n \pa$ is a cardinality 
of the set ${\cal K}_n$. The decoding function 
at the receiver is denoted by ${\psi}^{(n)}$. 
This function is formally defined by
$
{\psi}^{(n)}: {\cal Y}^{n} \to {\cal K}_n.
$
Let $c: {\cal X} \to [0,\infty)$ be a cost function.
The average cost on output of $\varphi^{(n)}$ 
must not exceed $\Gamma$. This condition is given by
$\varphi^{(n)}({\empty K}_n) \in {\cal S}_{\Gamma}^{(n)}$, where 
\beqno
{\cal S}_{\Gamma}^{(n)} 
&\defeq & \biggl\{{\Vcx} \in {\cal X}^n: 
\frac{1}{n}\sum_{t=1}^n c(x_t) \leq \Gamma 
\biggr\}.
\eeqno
The average error probabilities of decoding at the receiver 
is defined by 
\beqno
{\rm P}_{\rm e}^{(n)}
&=&{\rm P}_{\rm e}^{(n)}(\varphi^{(n)},\psi^{(n)}|W)
\defeq \Pr\{ \psi^{(n)}(Y^n)\neq K_n \}
\\
&=&1-\Pr\{\psi^{(n)}(Y^n)= K_n \}.
\eeqno
For $k\in {\cal K}_n$, set
$
{\cal D}(k)\defeq \{ {\Vcy} : \psi^{(n)}({\Vcy} )=k \}.
$
The families of sets 
$\{ {\cal D}(k) \}_{k\in {\cal K}_n}$ 
is called the decoding regions. Using the decoding region, 
${\rm P}_{\rm e}^{(n)}$ can be written as
\beqno
& &{\rm P}_{\rm e}^{(n)}={\rm P}_{\rm e}^{(n)}(\varphi^{(n)},\psi^{(n)}{}|W)
\\
&=&\frac{1}{|{\cal K}_n|} 
\sum_{k\in {\cal K}_n }
\sum_{\scs {\lvc y}  \notin {\cal D}(k)
      }
W^n\left( {\Vcy}   \left| \varphi^{(n)}(k) \right.\right).
\eeqno
Set
\beqno
& &{\rm P}^{(n)}_{\rm c}
  ={\rm P}^{(n)}_{\rm c}(\varphi^{(n)},\psi^{(n)}|W)
\defeq 
1-{\rm P}^{(n)}_{\rm e}(\varphi^{(n)}, \psi^{(n)} | W ).
\eeqno
The quantity ${\rm P}^{(n)}_{\rm c}$ is called the average 
correct probability of decoding. This quantity has 
the following form
\beqno
& &{\rm P}_{\rm c}^{(n)}
={\rm P}_{\rm c}^{(n)}(\varphi^{(n)},\psi^{(n)} | W )
\\
&=&\frac{1}{|{\cal K}_n|}\sum_{k\in {\cal K}_n }
\sum_{\scs  {\lvc y}  \in {\cal D}(k)}
W^n\left({\Vcy} \left| \varphi^{(n)}(k) \right.\right).
\eeqno
For given $\varepsilon$ $\in (0,1)$, 
$R$ is $\varepsilon$-{\it achievable} under $\Gamma$
if for any $\delta>0$, there exist a positive integer 
$n_0=n_0(\varepsilon,\delta)$ and a sequence of pairs 
$
\{(\varphi^{(n)},\psi^{(n)}): 
\varphi^{(n)}({\cal K}_n) \subseteq{\cal S}_{\Gamma}^{(n)} \}_{n=1}^{\infty}
$ 
such that for any $n\geq n_0(\varepsilon,\delta)$, 
\beqa 
{\rm P}_{{\rm e}}^{(n)}
(\varphi^{(n)},\psi^{(n)} | W )
&\leq &\varepsilon, 
\quad
\nbn \log \pa {\cal K}_n \pa  \geq  R-\delta.
\eeqa
The supremum of all $\varepsilon$-achievable $R$ under 
$\Gamma$ is denoted by ${C}_{\rm DMC}(\varepsilon,\Gamma|W)$.
We set
$$
C_{\rm DMC}(\Gamma|W)
\defeq \inf_{\varepsilon\in (0,1)}
C_{\rm DMC}(\varepsilon,\Gamma|W),
$$
which is called the channel capacity. The maximum error 
probability of decoding is defined by as follows:
\beqno
{\rm P}_{{\rm e,m}}^{(n)}&=&
   {\rm P}_{{\rm e,\irOlg{m}}}^{(n)}(\varphi^{(n)},\psi^{(n)}|W)
\\
& \defeq & \max_{k \in {\cal K}_n}
\Pr\{\psi^{(n)}({\Vcy} )\neq k|K_n=k\}.
\eeqno
Based on this quantity, we define 
${C}_{\rm DMC}(\varepsilon,\Gamma|W)$
by replacing ${\rm P}_{{\rm e}}^{(n)}(\varphi^{(n)},$ $\psi^{(n)}|W)$ 
in the definitions of $C_{\rm DMC}($ $\varepsilon,\Gamma|W)$ with 
${\rm P}_{{\rm e,m}}^{(n)}($ $\varphi^{(n)},\psi^{(n)}|W)$.
We set 
$$
{C}_{\irOlg{\rm m},\rm DMC}(\Gamma | \irBr{W})
=\inf_{\varepsilon\in (0,1)} 
C_{\irOlg{\rm m},\rm DMC}(\varepsilon,\Gamma |\irBr{W})
$$
which is called the \irOlg{maximum} capacity of the DMC.

We next define the channel capacities for finite length $n$.
For given $n$, a pair $(\varepsilon, R)$ is $n$-{\it achievable} 
under $\Gamma$
if there exists 
$(\varphi^{(n)},\psi^{(n)})$ with 
$\varphi^{(n)}({\cal K}_n) \subseteq {\cal S}_{\Gamma}^{(n)}$ 
such that 
\beqa 
{\rm P}_{{\rm e}}^{(n)}(\varphi^{(n)},\psi^{(n)}|W)
&\leq &\varepsilon, 
\quad
\nbn \log \pa {\cal K}_n \pa  \geq R.
\eeqa
We set
\beqno
& &{\cal R}_{\rm DMC}(n,\Gamma|W)
\\
& &\defeq
\{(\varepsilon, R): (\varepsilon, R)\mbox{ is }n\mbox{-{\it achievable}} 
\mbox{ under }\Gamma\}.
\eeqno 
Furthermore, set
\beqno
& &{C}_{\rm DMC}(n,\varepsilon, \Gamma |W)
\\
& &\defeq \max\{R:(\varepsilon, R)\in {\cal R}_{\rm DMC}(n,\Gamma|W)\},
\\
& &{\varepsilon}_{\rm DMC}(n,R,\Gamma |W)
\\
& &\defeq 
\min\{\varepsilon:(\varepsilon, R)\in {\cal R}_{\rm DMC}(n,\Gamma|W)\}.
\eeqno
We define ${\cal R}_{\rm m, DMC}(n,\varepsilon,\Gamma|W)$ 
by replacing ${\rm P}_{{\rm e}}^{(n)}(\varphi^{(n)},$ $\psi^{(n)}|W)$ 
in the definitions of ${\cal R}_{\rm DMC}($$n,\varepsilon,\Gamma|W)$ 
with ${\rm P}_{{\rm m,e}}^{(n)}($ $\varphi^{(n)},\psi^{(n)}|W)$.
We further define ${C}_{\rm m, DMC}(n,\varepsilon,\Gamma|W)$
and $\varepsilon_{\rm m, DMC}(n,\varepsilon,\Gamma|W)$
in a manner similar to the definitions in the case of average 
error criterion. Define 
\beqno
\underline{C}_{\rm DMC}(n,\varepsilon, \Gamma|W) 
&\defeq & 
\inf_{m\geq n} C_{\rm DMC}(m,\varepsilon, \Gamma|W),
\\
\underline{C}_{\rm m, DMC}(n,\varepsilon, \Gamma|W) 
&\defeq & 
\inf_{m\geq n} C_{\rm m, DMC}(m,\varepsilon, \Gamma|W).
\eeqno 
Then we have the following property. 
\begin{pr}\label{pr:pro0a} \ 
We have the following:
\beqno
C_{\rm DMC}(\varepsilon,\Gamma|W)
&=&\sup_{n\geq 1} 
\underline{C}_{\rm DMC}(n,\varepsilon,\Gamma|W),
\\
C_{\rm m, DMC}(\varepsilon,\Gamma|W)
&=&\sup_{n \geq 1}\underline{C}_{\rm m, DMC}(n,\varepsilon,\Gamma|W). 
\eeqno
\end{pr}

Proof of Property \ref{pr:pro0a} is given in Appendix \ref{sub:ApdaAAAA}. 
\newcommand{\ApdaAAAA}{
\subsection{
General Properties on 
${C}_{\rm DMC}(n,\varepsilon,\Gamma|W)$ 
and ${C}_{\rm m, DMC}($ $n,\varepsilon,\Gamma|W)$. 
}
\label{sub:ApdaAAAA}

In this appendix we prove Property \ref{pr:pro0a} describing 
general properties on ${C}_{\rm DMC}(n,\varepsilon,\Gamma|W)$ 
and ${C}_{\rm m, DMC}(n,\varepsilon,\Gamma|W)$. 

{\it Proof of Property \ref{pr:pro0a}:} We only prove the first equality 
of this property. A proof of the second equality is quite similar to that 
of the first equality. We omit the detail. 
We first prove the inequality 
$$
{C}_{\rm DMC}(\varepsilon,\Gamma|W)
\geq \sup_{m \geq 1} \underline{C}_{\rm DMC}(m,\varepsilon,\Gamma|W).
$$
We assume that 
$$
R \leq \sup_{m\geq 1} 
\underline{C}_{\rm DMC}(m,\varepsilon,\Gamma|W).$$ 
Then, there 
exists positive integer $m$ such that 
$R \leq \underline{C}_{\rm DMC}(m,$ $\varepsilon,\Gamma|W)$. 
Then, by the definition of 
$\underline{C}_{\rm DMC}(m,\varepsilon,\Gamma|W)$, we have that for any 
$n\geq m$, there exists a pair $(\varphi^{(n)},\psi^{(n)})$ with 
$\varphi^{(n)}({\cal K}_n) \subseteq {\cal S}_{\Gamma}^{(n)}$ 
such that 
\beq
{\rm P}_{{\rm e}}^{(n)}(\varphi^{(n)},\psi^{(n)}|W)\leq \varepsilon, 
\quad \frac{1}{n}\log |{\cal K}_n| \geq R.
\label{eqn: SssdF}
\eeq
It is obvious that under (\ref{eqn: SssdF}), we have 
for any $\delta>0$, and any $n\geq m $, we have 
\beq
{\rm P}_{{\rm e}}^{(n)}(\varphi^{(n)}, \psi^{(n)}|W)
\leq \varepsilon, 
\quad \frac{1}{n}\log |{\cal K}_n| \geq R-\delta.
\label{eqn: SssdFz}
\eeq
The bound (\ref{eqn: SssdFz}) implies that 
$R \leq C_{\rm DMC}(\varepsilon,\Gamma |W)$. Hence 
the bound 
$$
{C}_{\rm DMC}(\varepsilon,\Gamma|W)\geq 
\sup_{m \geq 1} \underline{C}_{\rm DMC}(m,\varepsilon,\Gamma|W)
$$ 
is proved. We next prove the reverse inequality. We assume that
$R\leq C_{\rm DMC}(\varepsilon,\Gamma |W)$. Then there exists 
$\{(\varphi^{(n)}, \psi^{(n)}): \varphi^{(n)}({\cal K}_n)$ 
$\subseteq {\cal S}_{\Gamma}^{(n)} \}_{n\geq 1}$ such that
for any $\delta> 0$ and any $n$ with 
$n \geq n_0=n_0(\varepsilon,\delta)$
we have that
\beq
{\rm P}_{{\rm e}}^{(n)}(\varphi^{(n)}, \psi^{(n)}|W)
\leq \varepsilon, 
\quad \frac{1}{n}\log |{\cal K}_n| \geq R-\delta.
\label{eqn: SssdFzz}
\eeq
The bound (\ref{eqn: SssdFzz}) implies that 
\beqno 
R-\delta &\leq & 
\underline{C}_{\rm DMC}(n_0,\varepsilon,\Gamma |W)
\leq \sup_{n \geq 1}\underline{C}_{\rm DMC}(n,\varepsilon,\Gamma |W).
\eeqno
On the other hand, by the first assumption we have 
$R-\delta\leq C_{\rm DMC}(\varepsilon,\Gamma |W)-\delta$. Hence, we have
$$
C_{\rm DMC}(\varepsilon,\Gamma |W)-\delta
\leq \sup_{n\geq 1}\underline{C}_{\rm DMC}(n,\varepsilon,\Gamma |W).
$$
Since we can take $\delta>0$ arbitrary small, we have
$$
C_{\rm DMC}(\varepsilon,\Gamma |W)
\leq \sup_{n\geq 1}\underline{C}_{\rm DMC}(n,\varepsilon,\Gamma |W),
$$ 
completing the proof. \hfill \IEEEQED
}

Set
\beq
C(\Gamma | W)=\max_{ \scs p_X \in {\cal P}({\cal X}):
          \atop{\scs
           {\rm E}_{p_X} c(X) \leq \Gamma 
          }
     }I(p_X,W), 
\eeq
where ${\cal P}({\cal X})$ is a set of probability distribution on 
${\cal X}$ and $I(p_X,W)$ stands for a mutual information 
between $X$ and $Y$ when input distribution of $X$ is $p_X$.
The following is a well known result.
\begin{Th}
\label{th:ddirect} 
{\rm For any DMC $W$, we have
$$
{C}_{\rm m, DMC}(\Gamma| W)={C}_{\rm DMC}(\Gamma|W)={C}(\Gamma| W).
$$
}
\end{Th}

Han \cite{Han98InfSpec} established the strong converse theorem for 
DMCs with input cost. His result is as follows. 

\begin{Th}[Han \cite{Han98InfSpec}] 
If $R>C(\Gamma | W)$, we have 
$$
\lim_{n\to\infty}{\rm P}_{\rm e}^{(n)}
(\varphi^{(n)}, \psi^{(n)}|W)=1
$$ 
for any 
$\{(\varphi^{(n)},\psi^{(n)}): 
\varphi^{(n)}({\cal K}_n) \subset {\cal S}_{\Gamma}^{(n)} \}_{n=1}^{\infty}
$ 
satisfying 
$$
\frac{1}{n}\liminf_{n\to \infty}\log M_n \geq R.
$$ 
\end{Th}

The following corollary immediately follows from this theorem. 

\begin{co}
For each fixed $\varepsilon$ $ \in (0,1)$ and any DMC 
$W$, we have 
$$
{C}_{\rm m, DMC}(\varepsilon, \Gamma |W)
={C}_{\rm DMC}(\varepsilon,\Gamma |W)={C}(\Gamma| W).
$$
\end{co}

\section{Main Results}

In this section we state our main results. We first define several quantities
necessary for describing those results. 
\begin{df}{\rm
For any $n$-sequence ${\Vcx} =x_{1}x_{2}\cdots $ 
$x_{n}\in {\calVarX}^{n}$, 
$n(x|{\Vcx} )$ denotes the number of $t$ such that $x_{t}=x$.  
The relative frequency $\left\{n(x|\varxun)/n\right\}_{x\in {\cal X}}$ 
of the components of ${\Vcx} $ is called the type of ${\Vcx} $ 
denoted by $P_{\lvc x}$. The set that consists of all 
the types on ${\cal X}$ is denoted by ${\cal P}_{n}({\cal X})$. 
Average cost for ${\vc x}\in {\cal X}^n$ is explicitly expressed 
with $P_{\lvc x}$. In fact we have the following: 
$$
\frac{1}{n}\sum_{t=1}^n c(x_t)=\sum_{x\in {\cal X}}c(x)P_{\lvc x}(x)
=\bar{c}_{P_{ {{\ssvc x}}}},
$$
where for $p \in {\cal P}({\cal X})$, we define
$$
\bar{c}_p\defeq \sum_{x\in {\cal X}}c(x)p(x).
$$}
\end{df}

\begin{df}
For any two $n$-sequences 
${\Vcx} =x_{1}$ $x_{2}$ $\cdots$ $x_{n}\in$ ${\cal X}^{n}$ and  
${\Vcy} =y_{1}$ $y_{2}$ $\cdots$ $y_{n}\in$ ${\cal Y}^{n}$, 
$n(x,y|{\Vcx} ,{\Vcy} )$ denotes the number 
of $t$ such that $(x_t,$ $y_t)=(x,$ $y)$.
The relative frequency 
$
\{n(x,y|{\Vcx} ,{\Vcy} )/n$ $\}_{(x,y)\in}$ ${}_{\calVarX\times\calVarY}
$ 
of the components of $({\Vcx} ,{\Vcy} )$ is called the joint type 
of $({\Vcx} ,{\Vcy} )$ denoted by $P_{{\lvc x} ,{\lvc y} }$. Furthermore, 
the set of all the joint type of $\calVarX\times\calVarY$ is denoted by 
${\cal P}_{n}(\calVarX\times\calVarY)$.
For each $({\Vcx} ,{\Vcy} )\in {\cal X}^n\times {\cal Y}^n$,
the joint type $P_{{\lvc x} ,{\lvc y} }$ induces the type $P_{{\lvc x} }$
given by
$$
P_{{\lvc x} }(x)=\sum_{y\in {\cal Y}}P_{{\lvc x} ,{\lvc y} }(x,y).
$$
Such type induced by a joint type is called the marginal type. 
For $P_{{\lvc x} }(x)>0$, we set 
$$ 
V_{{\lvc y} |{\lvc x} }(y|x)=
\frac{P_{{\lvc x} ,{\lvc y} }(x,y)}{P_{{\lvc x} }(x)}
=\frac{n(x,y|{\Vcx} ,{\Vcy} )}{n(x|{\Vcx} )}.
$$
For each $x\in {\cal X}$ with $P_{{\lvc x} }(x)>0$, 
$$
V_{{\lvc y} |{\lvc x} }(\cdot|x)=\{V_{{\lvc y} |{\lvc x} }(y|x)\}_{y\in {\cal Y}}
$$ 
becomes a conditional probability distribution. 
We call this the conditional 
type denoted by $V_{{\lvc y} |{\lvc x} }$. 
The formal definition of this quantity is 
$$
V_{{\lvc y} |{\lvc x} }=\{V(\cdot|x)\}_{x\in{\cal X}: P_{{\lvc x} }(x)>0}.
$$ 
\end{df}

\begin{df}
{\rm 
For $P\in {\cal P}_n({\cal X})$, let
${\cal V}_n({\cal Y}|P)$ be a set of all possible conditional type on 
${\cal Y}$ given $P$. 
Every $P_{{\lvc x} ,{\lvc y} }$ $\in {\cal P}({\cal X}\times{\cal Y})$ 
corresponds to $P_{{\lvc x} } \in {\cal P}_n({\cal X})$ and 
$V_{{\lvc y} |{\lvc x} }\in {\cal V}_n({\cal Y}|P_{{\lvc x} })$ 
in a one-to-one manner, that is, 
$$
P_{{\lvc x} ,{\lvc y} }=
(P_{{\lvc x} },V_{{\lvc y} |{\lvc x} }) \in 
\bigcup_{ P \in {\cal P}_n({\cal X})} \{ P \} \times {\cal V}_n({\cal Y}|P)
={\cal P}_n({\cal X} \times {\cal Y}).
$$ 
Conversely, for each $P\in {\cal P}_n({\cal X})$ and 
$V\in {\cal V}_n({\cal Y}| P)$, there exists $({\Vcx} ,{\Vcy} )
\in {\cal X}^n\times{\cal Y}^n$ such that
$$
P_{{\lvc x} ,{\lvc y} }=(P,V) \in 
\bigcup_{\tilde{P}\in {\cal P}_n({\cal X})}\{\tilde{P}\}
 \times {\cal V}_n({\cal Y}|\tilde{P})
={\cal P}_n({\cal X}\times{\cal Y}).
$$ 
For 
$
P_{{\lvc x} ,{\lvc y} }=(P,V) \in {\cal P}_n({\cal X}\times{\cal Y}),
$
the marginal type $P_{{\lvc y} }$ 
is induced by the product of $P$ and $V$, that is, 
$$
P_{{\lvc y} }(y)=\sum_{x\in {\cal X}}P(x)V(y|x). 
$$
We denote such $P_{{\lvc y} }$ by $PV$.
}\end{df}

Let $Y^n$ be an output 
from the noisy channel $W^n$ for the input 
$X^n=\varphi^{(n)}(K_n)$.
We have the following three propositions. Those are mathematical core 
of our main results. 
\begin{pro}\label{pro:pro1}
For any positive integer $n$, any $\gamma>0$, 
and any $(\varphi^{(n)},\psi^{(n)})$ with 
$\varphi^{(n)}({\cal K}_n)\subseteq {\cal S}_{\Gamma}^{(n)}$,
we have
\begin{align}
& {\rm P}_{\rm e}^{(n)}(\varphi^{(n)}, \psi^{(n)}|W)
\notag\\
&\geq {\rm Pr}\left\{
 \frac{1}{n}\log |{\cal K}_n| \geq 
\underline{I}(P_{\varphi^{(n)}(K_n)}, V_{Y^n|\varphi^{(n)}(K_n)}|W)
+\gamma\right\}
\notag \\
&\qquad -\nu_n(|{\cal Y}|)  
{\ExP}^{-n \gamma},
\label{eqn: SdaAFzz}
\end{align}
where 
$$
\nu_n=\nu_n(a)\defeq \binom{n+a-1}{a-1}\leq (n+1)^{a-1}, 
$$
$Y^n$ is an output from the noisy channel $\irBr{W}^n$ 
for the input $X^n=\varphi^{(n)}(K_n)$, 
and 
\begin{align*}
& \underline{I}(P_{\varphi^{(n)}(K_n)},V_{Y^n|\varphi^{(n)}(K_n)}|W)
\\
& \defeq I(P_{\varphi^{(n)}(K_n)},V_{Y^n|\varphi^{(n)}(K_n)})
\\
&\qquad -D(V_{Y^n|\varphi^{(n)}(K_n)}||W|P_{\varphi^{(n)}(K_n)}). 
\end{align*}
\end{pro}

\begin{pro}\label{pro:pro2}
For any positive interger $n$, 
    any $\gamma>0$, 
and any $P\in{\cal P}_n({\cal X})$ with 
$\bar{c}_P$ $\leq\Gamma$, there exists 
$(\varphi^{(n)},\psi^{(n)})$ with 
$\varphi^{(n)}({\cal K}_n)\subseteq T^n_P$ such that 
\begin{align}
&   {\rm P}_{\rm e}^{(n)}(\varphi^{(n)}, \psi^{(n)}|W)
\notag\\
&\leq 
{\rm Pr}\left\{
 \frac{1}{n}\log |{\cal K}_n| \geq 
J(P,V_{Y^n| \varphi^{(n)}(K_n)} |W)-\gamma\right\}
\notag \\
&\qquad 
+\kappa_n(|{\cal X}|){\ExP}^{-n \gamma}, 
\label{eqn: SdaAFzzz}
\end{align}
where 
$\kappa_n(a) \defeq {\rm e}^{\frac{a}{12}}(2\pi n)^{\frac{a-1}{2}}$
and  
\begin{align*}
& 
{J(\irBl{P, V_{Y^n |\varphi^{(n)}(K_n)}}|\irBr{W} )}
\notag \\
& \defeq \sum_{(x,y)\in {\cal X}\times{\cal Y}}
P(x) V_{Y^n |\varphi^{(n)}(K_n)}(y|x)\log \frac{W(y|x)}{(PW)(y)}.
\end{align*}
\end{pro}

\begin{pro}\label{pro:pro3}
For any positive integer $n$, any $\gamma>0$, and any $P\in{\cal P}_n({\cal X})$ with 
$\bar{c}_P$ $\leq\Gamma$, there exists $(\varphi^{(n)},\psi^{(n)})$ 
with $\varphi^{(n)}({\cal K}_n)\subseteq T^n_P$ such that
\begin{align}
&   {\rm P}_{\rm e}^{(n)}(\varphi^{(n)}, \psi^{(n)}|W)
\notag\\
&\leq 
{\rm Pr}\left\{
 \frac{1}{n}\log |{\cal K}_n| \geq 
I(P,V_{Y^n|\varphi^{(n)}(K_n)})-\gamma\right\}
\notag \\
&\qquad 
+\eta_n(|{\cal X}|,|{\cal Y}|){\ExP}^{-n \gamma}, 
\label{eqn: SdaAFzzzB}
\end{align}
where 
$\eta_n(a,b)\defeq \kappa_n({a})\nu_n(ab)$.
\end{pro}

Proofs of the above three propositions are given in the next section. 
To prove Proposition \ref{pro:pro1}, we introduce a new techinque for 
the meta converse lemma in the proof of converse coding 
theorems. On the other hand, proofs of Propositions \ref{pro:pro2}\ 
         and \ref{pro:pro3} are standard. 
Propositions \ref{pro:pro1}-\ref{pro:pro3}, together with 
a simple observation yield the following two theorems.
\begin{Th}\label{Th:Main}
\beqa
& & \max_{\gamma>0} \min_{\scs P \in {\cal P}_n({\cal X}):
\atop{\scs
\bar{c}_P \leq \Gamma 
}} 
\ba[t]{l}\left[
{\rm Pr }
\left\{
R \geq \underline{I}(P,V_{Y^n|X^n}|W)+\gamma \right.\right.
\vspace{1mm}\\
\qquad \left. \left| \left. P_{X^n}=P \right\} 
-\nu_n(|{\cal Y}|){\ExP}^{-n \gamma}\right. \right]
\ea
\nonumber\\
&\leq &\varepsilon_{\rm DMC}(n,R,\Gamma|W)
\notag\\
&\leq &
\min_{\gamma>\frac{1}{n}} \min_{\scs P \in {\cal P}_n({\cal X}):
\atop{\scs
\bar{c}_P \leq \Gamma 
}} 
\ba[t]{l}
\left[{\rm Pr}
\left\{
R \geq J(P,V_{Y^n|X^n}|W)-\gamma \right.\right.
\vspace{1mm}\\
\qquad \left. \left| \left. P_{X^n}=P \right\}+
2\kappa_n(|{\cal X}|){\ExP}^{-n \gamma}\right. \right].
\ea
\nonumber
\eeqa
\end{Th}

\begin{Th}\label{Th:MainB} 
\beqa
& & \max_{\gamma>0} \min_{\scs P \in {\cal P}_n({\cal X}):
\atop{\scs
\bar{c}_P \leq \Gamma 
}} 
\ba[t]{l}\left[
{\rm Pr }
\left\{
R \geq \underline{I}(P,V_{Y^n|X^n}|W)+\gamma \right.\right.
\vspace{1mm}\\
\qquad \left. \left| \left. P_{X^n}=P \right\} 
-\nu_n(|{\cal Y}|){\ExP}^{-n \gamma}\right. \right]
\ea
\nonumber\\
&\leq &\varepsilon_{\rm DMC}(n,R,\Gamma|W)
\notag\\
&\leq &
\min_{\gamma>\frac{1}{n}} \min_{\scs P \in {\cal P}_n({\cal X}):
\atop{\scs
\bar{c}_P \leq \Gamma 
}} 
\ba[t]{l}
\left[{\rm Pr}
\left\{
R \geq I(P,V_{Y^n|X^n})-\gamma \right.\right.
\vspace{1mm}\\
\qquad \left. \left| \left. P_{X^n}=P \right\}+
2\eta_n(|{\cal X}|,|{\cal Y}|)
{\ExP}^{-n \gamma}\right. \right].
\ea
\nonumber
\eeqa
\end{Th}

Proofs of Theorems \ref{Th:Main} and \ref{Th:MainB} will be given in 
Section \ref{sec:Secaa}.
\newcommand{\ProofThs}{
{\it Proof of Theorem \ref{Th:Main}:}
We first prove the first inequality of Theorem \ref{Th:Main}.
Let $(\varphi_{\rm opt}^{(n)},\psi_{\rm opt}^{(n)})$ 
be the optimal code that attains 
$\varepsilon_{\rm DMC}(n,R,\Gamma|W)$. By definitions we have the following:
\begin{align}
& \varepsilon_{\rm DMC}(n,R,\Gamma|W)
={\rm P}_{\rm e}^{(n)}(\varphi_{\rm opt}^{(n)}, 
                        \psi_{\rm opt}^{(n)}|W),
\label{eqn:aWss1}
\\
& \frac{1}{n}\log |{\cal K}_n|\geq R,
\label{eqn:aWss2}
\\
&\varphi^{(n)}({\cal K}_n)\subseteq {\cal S}_{\Gamma}^{(n)}.
\label{eqn:aWss3}
\end{align}
Set $X^n=\varphi_{\rm opt}^{(n)}(K_n)$. Note that $X^n$ is a uniformly 
distributed random variable with the cardinality $|{\cal K}_n|$ 
of the range of $X^n$. We also note that the condition (\ref{eqn:aWss3}) 
is equivalent to
\beq
X^n \in {\cal S}_{\Gamma}\Leftrightarrow 
\frac{1}{n}\sum_{t=1}^n c(X_t)=\bar{c}_{P_{X^n}} \leq \Gamma. 
\label{eqn:SxxFF}
\eeq
On a lower bound of $\varepsilon_{\rm DMC}(n,R,\Gamma|W)$, 
we have the following chain of inequalities:
\begin{align}
& \varepsilon_{\rm DMC}(n,R,\Gamma|W)
 \MEq{a}{\rm P}_{\rm e}^{(n)}(\varphi_{\rm opt}^{(n)}, 
                        \psi_{\rm opt}^{(n)}|W)
\notag\\
&\MGeq{b} {\rm Pr}\left\{
 \frac{1}{n}\log |{\cal K}_n| \geq 
\underline{I}(
P_{\varphi_{\rm opt}^{(n)}(K_n)},
V_{Y^n|\varphi_{\rm opt}^{(n)}(K_n)}|W)
+\gamma\right\}
\notag \\
&\qquad -\nu_n(|{\cal Y}|){\ExP}^{-n \gamma}
\notag\\
&\MGeq{c} {\rm Pr}\left\{
 R \geq \underline{I}(P_{X^n},
                      V_{Y^n|X^n}|W)
         +\gamma \right\}
-\nu_n(|{\cal Y}|){\ExP}^{-n \gamma}.
\label{eqn:AspQx}
\end{align}
Step (a) follows from (\ref{eqn:aWss1}).
Step (b) follows from Proposition \ref{pro:pro1}.
Step (c) follows from (\ref{eqn:aWss2}) and 
$X^n=\varphi_{\rm opt}^{(n)}(K_n)$.
On a lower bound of the first quantity in the right member of 
(\ref{eqn:AspQx}), we have the following chain of inequalities:
\begin{align}
&{\rm Pr}\left\{
 R \geq \underline{I}(P_{X^n},V_{Y^n|X^n}|W)+\gamma \right\}
\notag\\
&=\sum_{P\in {\cal P}({\cal X}^n)}
{\rm Pr}\left\{R \geq \underline{I}(P,V_{Y^n|X^n}|W)+\gamma 
\left| P_{X^n}=P \right. \right\} 
\notag\\
&\qquad \times {\rm Pr}\left\{P_{X^n}=P \right\}
\notag\\
&\MEq{a}
\sum_{\scs P \in {\cal P}({\cal X}^n):
      \atop{ \scs \bar{c}_P \leq \Gamma }}
{\rm Pr}\left\{R \geq \underline{I}(P,V_{Y^n|X^n}|W)+\gamma 
\left| P_{X^n}=P \right. \right\} 
\notag\\
&\qquad \times {\rm Pr} \left\{ P_{X^n}=P \right\}
\notag\\
&\geq 
\min_{\scs P \in {\cal P}_n({\cal X}):
\atop{\scs 
\bar{c}_P \leq \Gamma 
}} 
{\rm Pr}
\left\{R \geq \underline{I}(P,V_{Y^n|X^n}|W)+\gamma 
\left|  P_{X^n}=P \right. \right\} 
\notag\\
& \qquad \times 
\sum_{\scs P \in {\cal P}({\cal X}^n):
      \atop{ \scs \bar{c}_P \leq \Gamma }}
{\rm Pr}\left\{P_{X^n}=P \right\}
\notag\\
&\MEq{b} 
\min_{\scs P \in {\cal P}_n({\cal X}):
\atop{\scs 
\bar{c}_P \leq \Gamma 
}} 
{\rm Pr}
\left\{R \geq \underline{I}(P,V_{Y^n|X^n}|W)+\gamma 
\left|  P_{X^n}=P \right. \right\} 
\notag\\
& \qquad \times 
\sum_{\scs P \in {\cal P}({\cal X}^n)}
{\rm Pr}\left\{P_{X^n}=P \right\}
\notag\\
&=
\min_{\scs P \in {\cal P}_n({\cal X}):
\atop{\scs
\bar{c}_P \leq \Gamma 
}} 
{\rm Pr}
\left\{
R \geq \underline{I}(P,V_{Y^n|X^n}|W)+\gamma 
\left| P_{X^n}=P \right. \right\} 
\label{eqn:AzPz}
\end{align}
Steps (a) and (b) follow from (\ref{eqn:SxxFF}). 
From (\ref{eqn:AspQx}) and (\ref{eqn:AzPz}), we have 
\begin{align}
& \varepsilon_{\rm DMC}(n,R,\Gamma|W)
\nonumber\\
&\geq \min_{\scs P \in {\cal P}_n({\cal X}):
\atop{\scs
\bar{c}_P \leq \Gamma 
}} 
{\rm Pr }
\left\{R \geq \underline{I}(P,V_{Y^n|X^n}|W)+\gamma 
\left| P_{X^n}=P \right. \right\}   
\nonumber\\
& \qquad -\nu_n(|{\cal Y}|){\ExP}^{-n \gamma}.
\label{eqn:AsZcc}
\end{align}
Since (\ref{eqn:AsZcc}) holds for any $\gamma>0$, we have 
the first inequality of Theorem \ref{Th:Main}. 
We next prove the second inequality of Theorem \ref{Th:Main}.
We fix any positive interger $n$ and any positive $\gamma$. 
We choose $P^{\ast}\in {\cal P}({\cal X}^n)$ so that 
it attains the minimum of 
$$
{\rm Pr}
\left\{
R+\frac{1}{n} \geq J(P,V_{Y^n|X^n}|W)-\gamma \biggl| P_{X^n}=P \right\} 
$$
subject to $\bar{c}_P\leq \Gamma$.  
We choose $|{\cal K}_n|$ so that
$|{\cal K}_n|=2^{\lfloor nR \rfloor}$. 
By Proposition \ref{pro:pro2}, we have that 
for $P^* \in{\cal P}_n({\cal X})$ with 
$\bar{c}_{P^*}$ $\leq\Gamma$, 
there exists 
$(\varphi^{(n)},\psi^{(n)})$ with 
$\varphi^{(n)}({\cal K}_n)\subseteq T^n_{P^*}$ such that 
\begin{align}
&   {\rm P}_{\rm e}^{(n)}(\varphi^{(n)}, \psi^{(n)}|W)
\notag\\
&\leq 
{\rm Pr}\left\{
 \frac{1}{n}\log |{\cal K}_n| \geq 
J(P^*,V_{Y^n| \varphi^{(n)}(K_n)} |W)-\gamma\right\}
\notag \\
&\qquad 
+\kappa_n(|{\cal X}|){\ExP}^{-n \gamma}.
\label{eqn: SdaAzzzQ}
\end{align}
On an upper bound of $\varepsilon_{\rm DMC}(n,R,\Gamma|W)$, 
we have the following chain of inequalities:
\begin{align}
& \varepsilon_{\rm DMC}(n,R,\Gamma|W)
 \leq {\rm P}_{\rm e}^{(n)}(\varphi^{(n)}, 
                          \psi^{(n)}|W)
\notag\\
&\MLeq{a} {\rm Pr}\left\{
 \frac{1}{n}\log |{\cal K}_n| \geq 
J(P^*,V_{Y^n|\varphi^{(n)}(K_n)}|W)
-\gamma\right\}
\notag \\
&\qquad +\kappa_n(|{\cal X}|){\ExP}^{-n \gamma}
\notag\\
&\MLeq{b} {\rm Pr}\left\{
 R+\frac{1}{n} \geq J(P^*,V_{Y^n|X^n}|W)
         -\gamma \right\}
\notag \\
&\qquad 
+\kappa_n(|{\cal X}|){\ExP}^{-n\gamma}
\notag\\
&\MEq{c}
{\rm Pr}
\left\{
R+\frac{1}{n} \geq J(P^*,V_{Y^n|X^n}|W)-\gamma 
\biggl| P_{X^n}=P^* \right\}
\notag \\
&\qquad +\kappa_n(|{\cal X}|){\ExP}^{-n \gamma}
\notag \\
&\MEq{d}
\min_{\scs P \in {\cal P}_n({\cal X}):
\atop{\scs
\bar{c}_P \leq \Gamma 
}} 
{\rm Pr }
\left\{R \geq J(P,V_{Y^n|X^n}|W)-\gamma^{\prime} 
\left| P_{X^n}=P \right. \right\}   
\nonumber\\
& \qquad +2\kappa_n(|{\cal X}|){\ExP}^{-n \gamma^{\prime}}.
\label{eqn:AsZccQQQ}
\end{align}
Step (a) follows from (\ref{eqn: SdaAzzzQ}). 
Step (b) follows from that by $|{\cal K}_n|=2^{\lfloor n R \rfloor}$, 
we have $\log |{\cal K}_n|\leq nR +1$.
Step (c) follows from that $X^n \in T^n_{P^*}$.
Step (d) follows from the choice $\gamma^{\prime}=\gamma+(1/n)$.
Since (\ref{eqn:AsZccQQQ}) holds for any $\gamma^{\prime}>1/n$, we have 
the second inequality of Theorem \ref{Th:Main}. 
\hfill \IEEEQED 

{\it Proof of Theorem \ref{Th:MainB}:} 
The first inequality of Theorem \ref{Th:MainB} has already been proved.   
The second inequality can be proved by using Proposition \ref{pro:pro3}.
The proof is  similar to that of the second inequality. We omit the detail.
\hfill \IEEEQED 
}
By simple computations we can show that 
\beqa
& & \underline{I}(P,V_{Y^n|X^n}|\irBr{W})
\leq 
{J(\irBl{P,V_{Y^n|X^n}}|\irBr{W})}
\nonumber\\
& & \leq I(P,V_{Y^n|X^n}).
\label{eqn:asssZz}
\eeqa
By Theorem \ref{Th:Main}, we have the following result. 
\begin{co}
\beqa
& & \max_{\gamma>0} \min_{\scs P \in {\cal P}_n({\cal X}):
\atop{\scs
\bar{c}_P \leq \Gamma 
}} 
\ba[t]{l}\left[
{\rm Pr}
\left\{
\irb{R} \geq \irO{J(\irBl{P,V_{Y^n  |X^n  }}|\irBr{W} )}+\gamma \right.\right.
\vspace{1mm}\\
\qquad \quad\left. \left| \left. P_{X^n}=P \right\} 
-\nu_n(|{\cal Y}|){\ExP}^{-n \gamma}\right. \right]
\ea
\nonumber\\
&\leq &\irr{\varepsilon}_{\rm DMC}(n,\irb{R},\Gamma|\irBr{W})
\notag\\
&\leq &
\min_{\gamma> \frac{1}{n} } \min_{\scs P \in {\cal P}_n({\cal X}):
\atop{\scs
\bar{c}_P \leq \Gamma 
}} 
\ba[t]{l}
\left[{\rm Pr}
\left\{
\irb{R} \geq 
\irO{J(\irBl{P,V_{Y^n |X^n }}|\irBr{W})}-\gamma \right.\right.
\vspace{1mm}\\
\qquad \quad \left. \left| \left. P_{X^n}=P \right\}+
2\kappa_n(|{\cal X}|){\ExP}^{-n \gamma}\right. \right].
\ea
\nonumber
\eeqa
\end{co}

\section{Proofs of the Results}
\label{sec:Secaa}

In this section we give proofs of our main results. We first define 
several quantities and set necessary for the proofs. 
\begin{df}{
For $P \in {\cal P}_n({\cal X})$, set 
$
T^n_{P}\defeq \{{\Vcx}  :\,P_{{\lvc x} }=P \}.
$
For  
\beq
(P,V) \in \bigcup_{\tilde{P} \in {\cal P}_n({\cal X})} 
\{\tilde{P} \}\times {\cal V}_n({\cal Y}|\tilde{P})=
{\cal P}_n({\cal X} \times {\cal Y}),
\label{eqn:Asssgg}
\eeq
set 
$
T^n_{(P,V)}
\defeq \{({\Vcx},{\Vcy}) :\,P_{{\lvc x} ,{\lvc y} }=(P,V) \}.
$
For the above $(P,V)$, set 
$
T^n_{V}({\Vcx} ) 
\defeq \{{\Vcy} :\,P_{{\lvc x} ,{\lvc y} }=(P,V) \}.
$
}\end{df}

For set of types and joint types the following lemma holds. 
For the detail of the proof see Csisz\'ar and K\"orner\cite{ck}.
\begin{lm}\label{lm:Lem1}{\rm 
$\quad$
\begin{itemize}
\item[a)]
   $\ba[t]{l} 
   |{\cal P}_{n}({\calVarX})|=
   \nu_n(|{\cal X}|) \leq (n+1)^{|{\calVarX}|},\\  
   \mbox{For }P\in {\cal P}_{n}({\calVarX}),\\
   |{\cal V}_{n}({\calVarY}|P)| 
   =\nu_n(|{\cal X}||{\cal Y}|)
        \leq (n+1)^{|{\calVarX}||{\calVarY}|},\\  
   |{\cal P}_{n}({\calVarX}\times\calVarY)|
    =\nu_n(|{\cal X}||{\cal Y}|)\leq 
   (n+1)^{|{\calVarX}||{\calVarY}|}.
   \ea$
\item[b)] For $P\in {\cal P}_{n}(\calVarX)$ and 
    $V \in {\cal V}_{n}(\calVarY| P)$,  
\beqno
\hspace*{-4mm}[\kappa_n(|{\cal X}|)]^{-1}
{\ExP}^{nH(P)}&\leq&|T^n_{P}|\leq 2^{nH(P)},
\\
\hspace*{-4mm}
[\kappa_n(|{\cal Y}|)]^{-1}
{\ExP}^{nH(PV)}&\leq&|T^n_{PV}|\leq {\ExP}^{nH(PV)},
\eeqno
where $\kappa_n(a) 
\defeq {\rm e}^{\frac{a}{12}}(2\pi n)^{\frac{a-1}{2}}$.
For ${\Vcx}\in T^n_P$,
\beqno
\hspace*{-4mm}[\kappa_n(|{\cal X}||{\cal Y}|)]^{-1}
{\ExP}^{nH(V|P)}
&\leq&|T^n_{V}({\Vcx} )| \leq {\ExP}^{nH(V|P)},
\\
\hspace*{-4mm}[\kappa_n(|{\cal X}||{\cal Y}|)]^{-1}
{\ExP}^{nH(P,V)}
&\leq& |T^n_{(P,V)}|
\leq 
{\ExP}^{nH(P,V)}.
\eeqno
\item[c)] Suppose that $P\in {\cal P}_{n}(\calVarX)$, 
$V\in {\cal V}_{n}(\calVarY|P)$.
For ${\Vcx} \in T^n_{P}$ and ${\Vcy} \in T^n_{V}({\Vcx} )$,
\beqno
\hspace*{-6mm}p_{X}^n({\Vcx})
&=&{\ExP}^{-n[H(P)+D(P||p_X)]},
\\
\hspace*{-6mm}W^n({\Vcy}|{\Vcx})&=&{\ExP}^{-n[H(V|P)+D(V||W|P)]}.
\eeqno
\end{itemize}
}
\end{lm}

We first prove Proposition \ref{pro:pro1}. Set
\beqno
{\cal A}_l
&\defeq&
\{({\Vcx} , {\Vcy} ): 
W^n({\Vcy} |{\Vcx} )
\leq |{\cal K}_n| {\ExP}^{-n\gamma}
Q^{(l)}({\Vcy} )
\},
\\
{\cal A}_l({\Vcx} )&\defeq &
\{{\Vcy} :({\Vcx} ,{\Vcy} )\in {\cal A}_l\}. 
\eeqno
For ${\Vcx} \in {\cal X}^n$ and $V\in {\cal V}_n(P_{{\lvc x} })$, 
we set 
\beqno
& &{\cal A}_{l,V}({\Vcx} )
\defeq {\cal A}_l({\Vcx} )\cap T^n_V({\Vcx} )
\\
&=&
\{{\Vcy}  \in T^n_V({\Vcx} ) : 
W^n({\Vcy} |{\Vcx} )
\leq  |{\cal K}_n| {\ExP}^{-n\gamma}
Q^{(l)}({\Vcy} )\}.
\eeqno
The following lemma is useful for the proof.
\begin{lm}\label{lm:Ohzzz}
For any $\gamma>0$ and for any $(\varphi^{(n)},\psi^{(n)})$, 
we have
\begin{align}
& {\rm P}_{\rm e}^{(n)}
(\varphi^{(n)},\psi^{(n)}{}|W)
\geq {\rm Pr}
\hugel \bigcup_{l=1}^{L} \biggl[ 
\nonumber\\
& \left.
\frac{1}{n}\log |{\cal K}_n| 
\geq \frac{1}{n}\log
\frac{W^n(Y^n|X^n)}{Q^{(l)}(Y^n)}+\gamma\right]\huger
-L{\ExP}^{-n\gamma}. 
\label{eqn:azsad00}
\end{align}
In (\ref{eqn:azsad00}) we can choose any probability 
distribution $Q^{(l)},l=$ $1,2,\cdots, L$ on ${\cal Y}^n$. 
\end{lm}

\newcommand{\Apda}{
\subsection{
Proof of Lemmas \ref{lm:Ohzzz} and \ref{lm:FbOhzzz} 
}\label{sub:Apda}

In this appendix we prove Lemmas \ref{lm:Ohzzz} 
and \ref{lm:FbOhzzz}. 
}{

{\it Proof:} The bound (\ref{eqn:azsad00}) we wish to show 
is equivalent to 
\begin{align*}
& 1-{\rm P}_{\rm e}^{(n)}
(\varphi^{(n)},\psi^{(n)}{}|W)
\leq {\rm Pr}
\hugel
\nonumber\\
& 
\bigcap_{l=1}^{L}\left[ 
\frac{1}{n}\log |{\cal K}_n| 
< \frac{1}{n}\log
\frac{W^n(Y^n|X^n)}{Q^{(l)}(Y^n)}+\gamma\right]\huger
+L{\ExP}^{-n\gamma}.
\end{align*}
In the following argument we prove this bound. 
Then we have the following: 
\beqno
& &1-{\rm P}_{\rm e}^{(n)}
(\varphi^{(n)},\psi^{(n)}{}|W)
\\
&=&\frac{1}{|{\cal K}_n| } 
\sum_{k\in {\cal K}_n }
\sum_{ \scs {\lvc y} \in
          {\cal D}(k) 
              \atop{\scs
               \bigcap \left[\bigcap_{l=1}^L {\cal A}_l^c(\varphi^{(n)}(k))\right]} 
   }  
W^n({\Vcy} |\varphi^{(n)}(k))
\\
&&+\frac{1}{|{\cal K}_n| } 
  \sum_{k\in {\cal K}_n }
  \sum_{\scs {\lvc y} \in 
          {\cal D}(k) 
              \atop{ \scs 
              \bigcap  \left[\bigcap_{l=1}^L {\cal A}_l^c(\varphi^{(n)}(k))\right]^c}
    }  
W^n({\Vcy} |\varphi^{(n)}(k))
\\
&\leq& \Delta_0+\sum_{l=1}^L \Delta_l,
\eeqno
where
\beqno
\Delta_0
&\defeq &\frac{1}{|{\cal K}_n|}  
\sum_{k\in {\cal K}_n }
\sum_{\scs {\lvc y} \in 
\bigcap_{l=1}^L {\cal A}_l^{c}(\varphi^{(n)}(k))
}
W^n({\Vcy} |\varphi^{(n)}(k)),
\\
\Delta_l
&\defeq &
  \frac{1}{|{\cal K}_n| } 
  \sum_{k\in {\cal K}_n }
  \sum_{\scs {\lvc y} \in {\cal D}(k) \cap {\cal A}_l(\varphi^{(n)}(k)) 
       \atop{\scs 
       }
    }  
W^n({\Vcy} |\varphi^{(n)}(k)).
\eeqno
On the quantity $\Delta_0$, we have 
\beqno
\Delta_0&=
&
{\rm Pr}\hugel
\bigcap_{l=1}^{L}\left[ 
\frac{1}{n}\log |{\cal K}_n| 
< \frac{1}{n}\log
\frac{W^n(Y^n|X^n)}{Q^{(l)}(Y^n)}+\gamma\right]\huger.
\eeqno
Hence it suffices to show $\Delta_l\leq {\ExP}^{-n\gamma}$ for 
$l=1,2,\cdots, L$ to prove Lemma \ref{lm:Ohzzz}. 
We have the following chain of inequalities: 
\begin{align*}
&\Delta_l
=  \frac{1}{|{\cal K}_n| } \sum_{k\in {\cal K}_n }
  \sum_{\scs 
         \atop{\scs 
             {\lvc y}  \in {\cal D}(k): 
                \atop{\scs 
                    W^n({\lvc y} |\varphi^{(n)}(k)) 
                    \atop{\scs   
                      \leq \ExP^{-n\gamma}|{\cal K}_n| Q^{(l)}({\lvc y} )
                    }
                }
         }
    }W^n({\Vcy} |\varphi^{(n)}(k)) 
\\
&\leq 
  {\ExP}^{-n\gamma}
  \sum_{k\in {\cal K}_n }
  \sum_{\scs
           \atop {\scs 
            {\lvc y}  \in {\cal D}(k) 
       }
    }
Q^{(l)}({\Vcy} ) 
= {\ExP}^{-n\gamma} 
   \sum_{ k\in {\cal K}_n }
   Q^{(l)}\left({\cal D}(k) \right) 
\\
&={\ExP}^{-n\gamma} 
  Q^{(l)}\left( \bigcup_{k\in {\cal K}_n}{\cal D}(k)\right) 
\leq {\ExP}^{-n\gamma}.
\end{align*}
Thus Lemma \ref{lm:Ohzzz} is proved. 
\hfill\IEEEQED
}%

{\it Proof of Proposition \ref{pro:pro1}:} 
We set 
\begin{align}
&B\defeq  
{\rm Pr}
\hugel
\bigcup_{l=1}^{L}\left[ 
\frac{1}{n}\log |{\cal K}_n| 
\geq \frac{1}{n}\log
\frac{W^n(Y^n|X^n)}{Q^{(l)}(Y^n)}+\gamma\right]\huger
\nonumber\\
&=\frac{1}{|{\cal K}_n|}
\sum_{k\in {\cal K}_n }\sum_{\scs {\lvc y} \in 
\bigcup_{l=1}^L {\cal A}_l(\varphi^{(n)}(k))}
W^n({\Vcy} |\varphi^{(n)}(k))
\nonumber\\
&=\frac{1}{|{\cal K}_n|}\sum_{k\in {\cal K}_n}B(k),
\nonumber
\end{align}
where
\beqno
B(k)\defeq 
\sum_{
\scs {\lvc y}  \in 
\bigcup_{l=1}^L {\cal A}_l(\varphi^{(n)}(k))
}
W^n({\Vcy} |\varphi^{(n)}(k)).
\eeqno
For 
$(P,V)\in {\cal P}_n({\cal X})\times {\cal V}_n({\cal Y}|P)$, define 
$\iota_1$ by
$
\iota_1(P,V)=PV \in {\cal P}_n({\cal Y}). 
$
Let $L=|{\cal P}_n({\cal Y})|$ and let 
$
\iota_2:{\cal P}_n({\cal Y}) \to \{1,2,\cdots, L \}
$ 
be a one-to-one mapping. 
Using $\iota_1$ and $\iota_2$, we define the map $l$ by 
$l=\iota_2 \circ \iota_1$, i.e., for 
$(P,V)\in {\cal P}_n($ ${\cal X})\times{\cal V}_n({\cal Y}|P)$,
define 
$$
l=\iota_2\circ\iota_1(P,V)=\iota_2(PV) \in \{1,\cdots,L\}. 
$$
For each $l=1,2,\cdots, L$, we choose $Q^{(l)}$ so that
it is the uniform distribution over $T^n_{\iota_2^{-1}(l)}
=T^n_{PV}$ for $\iota_2(PV)=l$, i.e.,
$$
Q^{(l)}({\Vcy} )=
\left\{
 \ba[c]{cl} 
\ds \frac{1}{|T^n_{PV}|}&\mbox{ if }{\Vcy} \in  {T^n_{PV}},
\vspace*{1mm}\\
0 &\mbox{ otherwise.} 
\ea
\right.
$$ 
For each $k\in{\cal K}_n$, we have the following chain of inequalities: 
\beqa
& &B(k)
=\sum_{
\scs {\lvc y}  \in 
\bigcup_{l=1}^L {\cal A}_l(\varphi^{(n)}(k))
}
W^n({\Vcy} |\varphi^{(n)}(k))
\nonumber\\
&=&
\sum_{V\in {\cal V}_n\left({\cal Y}\left| P_{\varphi^{(n)}(k)}\right. \right)}
\sum_{
\scs {\lvc y}  \in 
\bigcup_{l=1}^L {\cal A}_{l,V}(\varphi^{(n)}(k))}
W^n({\Vcy} |\varphi^{(n)}(k))
\nonumber\\
&\geq&
\sum_{ V \in {\cal V}_n\left({\cal Y}\left| P_{\varphi^{(n)}(k)}\right. \right)}
\sum_{
\scs {\lvc y}  \in 
{\cal A}_{l^*
,V}(\varphi^{(n)}(k))}
W^n({\Vcy} |\varphi^{(n)}(k)),
\quad
\label{eqn:Asxx}
\eeqa
where the quantity $l^\ast$ in the last step is the index so that
\beqno
& &
l^*=l \left( P_{\varphi^{(n)}(k)},V \right)
=l \left( P_{\varphi^{(n)}(k)}, V_{{\lvc y} |\varphi^{(n)}(k)}\right), 
\nonumber\\
& &\mbox{for }{\Vcy}  \in T^n_{V}(\varphi^{(n)}(k)).
\eeqno
Note that
\begin{align}
& {\cal A}_{l^*,V}(\varphi^{(n)}(k))
=
\Bigl\{{\Vcy}  \in T^n_V(\varphi^{(n)}(k)): 
\nonumber\\
& W^n({\Vcy} |\varphi^{(n)}(k))
\leq |{\cal K}_n| {\rm 2}^{-n\gamma}
Q^{(l^*)}({\Vcy} )\Bigr\}
\nonumber\\
&=\left\{ \Vcy \in T^n_V(\varphi^{(n)}(k)): 
W^n({\Vcy}|\varphi^{(n)}(k)) \leq  
\frac{|{\cal K}_n| {\ExP}^{-n\gamma}}
{\left| T^n_{P_{\varphi^{(n)}(k)}V} \right| }
\right\}
\nonumber\\
&=\hugel {\Vcy}  \in T^n_V(\varphi^{(n)}(k)):
\nonumber\\
& \quad |{\cal K}_n| 
\geq W^n({\Vcy} |\varphi^{(n)}(k)) 
\left|T^n_{P_{\varphi^{(n)}(k)}V}\right|
{\rm 2}^{n\gamma} \huger.
\label{eqn:aSddZa}
\end{align}
By Lemma \ref{lm:Lem1} parts b) and c), we have 
\beqno
& &\left|T^n_{P_{\varphi^{(n)}(k)}V}\right|
 \leq {\ExP}^{n H\left(P_{\varphi^{(n)}(k)}V\right)}, 
\\
& &W^n({\Vcy} |\varphi^{(n)}(k))
=
{\ExP}^{-n\left[H\left(V \left| P_{\varphi^{(n)}(k)}\right.\right)
+D\left(V\left|\left|W \left| P_{\varphi^{(n)}(k)}\right.\right.
\right)\right.\right]}. 
\eeqno
Using those bounds, we obtain
\beqa
& & \frac{1}{n} \log |{\cal K}_n|
\geq \underline{I}( P_{\varphi^{(n)}(k)},V | W )+\gamma
\nonumber\\
&\Rightarrow&
|{\cal K}_n| 
\geq W^n({\Vcy} |\varphi^{(n)}(k)) \left|T^n_{P_{\varphi^{(n)}(k)}V}\right|
{\rm 2}^{n\gamma}.
\label{eqn:aSddZ}
\eeqa
From (\ref{eqn:aSddZa}) and (\ref{eqn:aSddZ}), we obtain
\beqa
& &{\cal A}_{l^*,V}(\varphi^{(n)}(k))
 \supseteq \hugel {\Vcy}  \in T^n_V(\varphi^{(n)}(k)):
\nonumber\\
& &\quad \frac{1}{n} \log |{\cal K}_n|
\geq \underline{I}( P_{\varphi^{(n)}(k)},V | W )+\gamma \huger
\nonumber\\
&=&\hugel {\Vcy}  \in T^n_V(\varphi^{(n)}(k)): \frac{1}{n} \log |{\cal K}_n|
\nonumber\\
& &\quad 
\geq \underline{I}( P_{\varphi^{(n)}(k)},V_{{\lvc y} |\varphi^{(n)}(k)}
|W)+\gamma \huger.
\label{eqn:Sdff}
\eeqa
From (\ref{eqn:Asxx}) and (\ref{eqn:Sdff}), we have
\begin{align}
& B(k)
\geq 
{\rm Pr}\hugel \frac{1}{n} \log |{\cal K}_n|
\geq 
\underline{I}( P_{\varphi^{(n)}(k)},V_{Y^n|\varphi^{(n)}(k)}|W)
\nonumber\\
& \qquad\qquad\quad\: 
+\gamma \huger \mbox{ for }k\in {\cal K}_n.
\label{eqn:SdffX}
\end{align}
Combining all results we have obtained so far, we have
\begin{align}
& {\rm P}_{\rm e}^{(n)}(\varphi^{(n)},\psi^{(n)}|W)
\MGeq{a} B - L {\ExP}^{-n \gamma}
\nonumber\\
&\MEq{b}  \frac{1}{|{\cal K}_n|}
\sum_{ k\in {\cal K}_n}B(k) - |{\cal P}_n({\cal Y})|{\ExP}^{-n \gamma}
\nonumber\\
& \MEq{c}  \frac{1}{|{\cal K}_n|}
\sum_{ k\in {\cal K}_n}B(k) - \nu_n({|{\cal Y}|}){\ExP}^{-n \gamma}
\nonumber\\
&\MGeq{d}  
{\rm Pr}
\left\{
\frac{1}{n} \log |{\cal K}_n| 
\geq \underline{I}( P_{\varphi^{(n)}(K_n)},V_{Y^n|\varphi^{(n)}(K_n)}|W)
+\gamma 
\right\}
\nonumber\\
&\quad -\nu_n({|{\cal Y}|}){\ExP}^{-n \gamma}.
\nonumber
\end{align}
Step (a) follows from Lemma \ref{lm:Ohzzz}. 
Step (b) follows from the choice $L=|{\cal P}_n({\cal Y})|$.
Step (c) follows from Lemma \ref{lm:Lem1} part a).
Step (d) follows from (\ref{eqn:SdffX}).
\hfill \IEEEQED

We next prove Proposition \ref{pro:pro2}.
Using an argument of random coding,  
we show an exsitance of encoding and decoding schemes 
to attain the upper bound of 
${\rm P}_{\rm e}^{(n)}(\varphi^{(n)},\psi^{(n)}|W)$. 

\noindent
\underline{Random Coding:} Fix $P\in {\cal P}_n({\cal X})$ such that
$\bar{c}_P\leq \Gamma$. For each $k\in {\cal K}_n$, we generate 
${\Vcx} ={\Vcx} (k)$ according to the uniform distribution 
over $T^n_P$. 

\noindent
\underline{Encoding:} For each $k \in {\cal K}_n$, we 
define $\varphi^{(n)}(k)$ by $\varphi^{(n)}(k)={\Vcx} (k)$.

\noindent
\underline{Decoding:} 
Define 
$$
J({\Vcx};{\Vcy} |W)\defeq J(P_{\lvc x}, V_{{\lvc y}|{\lvc x}}|W).
$$
Set 
\beqno
{\cal T}_{\gamma}^{(n)}
&\defeq&
\Bigl\{({\Vcx} , {\Vcy} )\in {\cal X}^n\times {\cal Y}^n:
\\
& & \qquad J({\Vcx} ;{\Vcy}|W)
\geq \frac{1}{n}\log |{\cal K}_n| +\gamma \Bigr\},
\\
{\cal T}_{\gamma,1}^{(n)}
&\defeq& \left\{{\Vcx} : ({\Vcx} , {\Vcy} )\in {\cal T}_{\gamma}^{(n)}
\mbox{ for some }{\Vcy}  \in {\cal X}^n\right\},
\\
{\cal T}_{\gamma,2}^{(n)}
&\defeq& \left\{{\Vcy} : ({\Vcx} , {\Vcy} )\in {\cal T}_{\gamma}^{(n)}
\mbox{ for some }{\Vcx}  \in {\cal X}^n\right\}.
\eeqno
For ${\Vcy}  \in {\cal T}_{\gamma,2}^{(n)}$, we set 
\beqno
{\cal T}_{\gamma,1}^{(n)}({\Vcy} )
\defeq \left\{{\Vcx} : ({\Vcx} ,{\Vcy} )\in {\cal T}_{\gamma}\right\}.
\eeqno
Similarly, for ${\Vcx}  \in {\cal T}_{\gamma,1}^{(n)}$, we set 
\beqno
{\cal T}_{\gamma,2}^{(n)}({\Vcx} )
\defeq \left\{{\Vcx} : ({\Vcx} ,{\Vcy} )\in {\cal T}_{\gamma}\right\}.
\eeqno
For received sequence ${\Vcy} \in {\cal Y}^n$, we 
define the decoder function by
\begin{align}
& \psi^{(n)}({\Vcy} )
\notag\\
&\defeq \left\{
\ba{cl}
\hat{k} 
&\mbox{ if } {\Vcx} (\hat{k})\in {\cal T}_{\gamma,1}^{(n)}({\Vcy} )
\mbox{ and } {\Vcx} (\tilde{k}) 
\notin {\cal T}_{\gamma,1}^{(n)}({\Vcy} )
\\
&\mbox{ for all } \tilde{k}\in {\cal K}_n-\{\hat{k}\},
\\
\mbox{0}&\mbox{ otherwise.}
\ea
\right.
\end{align}

\noindent
\underline{Error Probability Analysis:} 
For $P\in {\cal P}_n({\cal X})$, define 
a probability distribution $Q$ on ${\cal Y}^n$ 
by
$$
Q({\Vcy} ) \defeq \frac{1}{|T^n_P|}
\sum_{{\lvc x}  \in T^n_P}W^n({\Vcy} |{\Vcx} ). 
$$
Then we have the following lemma.
\begin{lm}\label{lm:LemmaA} 
Fix $(P,V)$ $\in {\cal P}_n ({\cal X} \times {\cal Y})$ arbitrary.
Let $PV\in {\cal P}_n({\cal Y})$ be a type on ${\cal Y}$ 
induced by $(P,V)$. Then for any 
${\Vcy} \in T^n_{PV}$, we have 
\beqa
Q({\Vcy} )\leq 
\kappa_n(|{\cal X}|)2^{-n[H(PV)+D(PV||PW)]}. 
\label{eqn:Asdff}
\eeqa
Furthermore, for any $({\Vcx} , {\Vcy} ) \in T_{(P,V)}^n$, we have 
\beqa
\frac{Q({\Vcy} )}{W^n({\Vcy} |{\Vcx} )}
&\leq & \kappa_n(|{\cal X}|)2^{-n J(P,V|W)}
\nonumber\\
&=&
\kappa_n(|{\cal X}|)2^{-n J ({\lvc x} ;{\lvc y} |W)}.
\label{eqn:Asdffzz}
\eeqa
\end{lm}

{\it Proof:} We first prove (\ref{eqn:Asdff}). We have the following 
chain of inequalities:
\begin{align}
& Q({\Vcy} ) = \frac{1}{|T^n_P|}
\sum_{{\lvc x}  \in T^n_P}W^n({\Vcy} |{\Vcx} )
\notag\\
&
\MLeq{a}
\sum_{{\lvc x}  \in T^n_P}W^n({\Vcy} |{\Vcx} )
\kappa_n(|{\cal X}|)2^{-nH(P)}  
\notag\\
&=
\kappa_n(|{\cal X}|)
\sum_{{\lvc x}  \in T^n_P}
\prod_{t=1}^n\left[W(y_t|x_t) P({x}_t)\right]
\notag\\
&\MLeq{b}
\kappa_n(|{\cal X}|)
\sum_{{\lvc x}  \in {\cal X}^n}
\prod_{t=1}^n\left[W(y_t|x_t)P({x}_t)\right]  
\notag\\
&=\kappa_n(|{\cal X}|)
\prod_{t=1}^n\left[\sum_{x_t \in {\cal X}}
W(y_t|x_t)P({x}_t)\right]  
\notag\\
& =\kappa_n(|{\cal X}|)
\prod_{t=1}^n(PW)(y_t)
\notag\\
& \MEq{b}\kappa_n(|{\cal X}|)2^{-n[H(PV)+D(PV||PW)]}.  
\notag
\end{align}
Step (a) follows from Lemma \ref{lm:Lem1} part b). 
Step (b) follows from Lemma \ref{lm:Lem1} part c). 
We next prove (\ref{eqn:Asdffzz}). 
When $({\Vcx} ,{\Vcy} )\in T^n_{(P,V)}$ we have 
\beqa
W^n({\Vcy} |{\Vcx} )=2^{-n[H(P|V)+D(V||W|P)]}.
\label{eqn:AsszPp}
\eeqa
From (\ref{eqn:Asdff}) and (\ref{eqn:AsszPp}), we have 
\begin{align*}
& \frac{Q({\Vcy} )}{W^n({\Vcy} |{\Vcx} )}
\leq \frac{\kappa_n(|{\cal X}|)2^{-n[H(PV)+D(PV||PW)}]}
{2^{-n[H(P|V)+D(V||W|P)]}}
\nonumber\\
&=\kappa_n(|{\cal X}|)2^{-n J(P,V|W)}=
\kappa_n(|{\cal X}|)2^{-nJ({\lvc x} ;{\lvc y} |W)}.
\end{align*}
Thus (\ref{eqn:Asdffzz}) is proved. 
\hfill\IEEEQED 

In the following argument we let ${\sf P}$ denote a probability 
measure based on the randomness of the choice 
of $\{{\Vcx} (k)\}_{k\in {\cal K}_n}$. 
Let ${\sf E}$ denote an expectation based on the 
randomness of the choice of $\{{\Vcx} (k)\}_{k\in {\cal K}_n}$. 

{\it Proof of Proposition \ref{pro:pro2}:} 
We use a pair of proposed encoder and decoder functions denoted by 
$(\varphi^{(n)},\psi^{(n)})$. By the construction of $\varphi^{(n)}$, 
we have that for any $k\in {\cal K}_n$, ${\Vcx} (k) \in T^n_P.$
Fix $(P,V)$ $\in {\cal P}_n ({\cal X} \times {\cal Y})$
arbitrary. For ${\Vcx} \in T_P^n$, set
\beqno
& &{\cal T}_{\gamma,2,V}^{(n)}({\Vcx} )
\\
&\defeq& \left\{ {\Vcy}  \in T_V^n({\Vcx} ):
J(P,V|W)\geq \frac 1 n \log |{\cal K}_n|+\gamma \right\}
\\
&=& \left\{ {\Vcy}  \in T_V^n({\Vcx} ):
J({\Vcx} ,{\Vcy} |W)\geq  \frac 1 n 
\log |{\cal K}_n|+\gamma \right\}.
\eeqno
Furthermore, set
\beqno
& &\breve{\cal T}_{\gamma,2,V}^{(n)}({\Vcx} )
\\
&&\defeq \left\{ {\Vcy}  \in T_V^n({\Vcx} ):
J(P,V|W) < \frac 1 n \log |{\cal K}_n|+\gamma \right\}.
\eeqno
By definition it is obvious that for every ${\Vcx} \in T_P^n$, 
\beqno
&&{\cal T}_{\gamma,2,V}^{(n)}({\Vcx})
\cap \breve{\cal T}_{\gamma,2,V}^{(n)}({\Vcx})=\emptyset,
\\
&&{\cal T}_{\gamma,2,V}^{(n)}({\Vcx})
\cup \breve{\cal T}_{\gamma,2,V}^{(n)}({\Vcx})=T_V^n({\Vcx} ).
\eeqno
For $k\in {\cal K}_n$ and 
for $({\Vcx} (k),{\Vcy} )\in {\cal X}^n\times {\cal Y}^n$, 
define
$$
\chi_{{\lvc y} |{\lvc x} (k)}(\varphi^{(n)},\psi^{(n)})
\defeq\left\{ 
\ba{ll} 1 &\mbox{ if }\psi^{(n)}({\Vcy} )\neq k,
\\
 0 & \mbox{ otherwise. }
\ea
\right.
$$
Then we have 
${\rm P}_{\rm e}^{(n)}(\varphi^{(n)},\psi^{(n)}|W)
\newcommand{\omitA}{
&=&\frac{1}{|{\cal K}_n|}
\sum_{k\in {\cal K}_n}
\sum_{{\lvc y} \in {\cal Y}^n}
\chi_{{\lvc y} |{\Vcx} ({k})}(\varphi^{(n)},\psi^{(n)})
W^n({\Vcy} |{\Vcx} (k))
\\
&=&
\frac{1}{|{\cal K}_n|}
\sum_{k\in {\cal K}_n}
\sum_{V \in {\cal V}_n({\cal Y}|P_{{\Vcx} (k)})}
\sum_{{\lvc y}  \in T_V^n({\Vcx} )}1
\\
& &\qquad \times \chi_{{\lvc y} |{\Vcx} ({k})}(\varphi^{(n)},\psi^{(n)})
W^n({\Vcy} |{\Vcx} (k))
\\
&=&\Lambda_1+\Lambda_2, 
}
=\Lambda_1+\Lambda_2, 
$
where
\beqno
\Lambda_1&=&\frac{1}{|{\cal K}_n|}
\sum_{k \in {\cal K}_n}
\sum_{V \in {\cal V}_n({\cal Y}|P)}
\sum_{{\lvc y}  \in \breve{\cal T}_{\gamma,2,V}^{(n)}({\Vcx} )}1
\\
& &\qquad\times 
\chi_{{\lvc y} |{\Vcx} ({k})}(\varphi^{(n)},
\psi^{(n)})W^n({\Vcy} |{\Vcx} (k)),
\\
\Lambda_2&=&\frac{1}{|{\cal K}_n|}
\sum_{k \in {\cal K}_n}
\sum_{V \in {\cal V}_n({\cal Y}|P)}
\sum_{{\lvc y}  \in {\cal T}_{\gamma,2,V}^{(n)}({\lvc x} )}1
\\
& &\qquad \times 
\chi_{{\lvc y} |{\Vcx} ({k})}(\varphi^{(n)},\psi^{(n)})W^n({\Vcy} |{\Vcx} (k)).
\eeqno
On an upper bound of $\Lambda_1$, we have the following:
\beqno
\Lambda_1
& \leq & \frac{1}{|{\cal K}_n|}
\sum_{k \in {\cal K}_n}
\sum_{V \in {\cal V}_n({\cal Y}|P)}
\sum_{{\lvc y}  \in \breve{\cal T}_{\gamma,2,V}^{(n)}({\Vcx} )}
W^n({\Vcy} |{\Vcx} (k))
\\
&=&
{\rm Pr}\left\{
 \frac{1}{n}\log |{\cal K}_n| \geq 
J(P,V_{Y^n|\varphi^{(n)}(K_n)}|W)-\gamma\right\}.
\eeqno 
We next derive an upper bound of $\Lambda_2$. On this bound 
we have the following chain of inequalities:  
\begin{align}
& \Lambda_2
=\frac{1}{|{\cal K}_n|}
\sum_{k \in {\cal K}_n}
\sum_{V \in {\cal V}_n({\cal Y}|P)}
\sum_{\scs {\lvc y}  \in 
\left[\scs \bigcup_{m \neq k}{\cal T}_{\gamma,2,V}^{(n)}({\Vcx} (m))\right]
      \atop{\scs 
      \bigcap {\cal T}_{\gamma,2,V}^{(n)}({\Vcx} (k))
      }
}1
\notag\\
& \qquad \times 
\chi_{{\lvc y} |{\Vcx} ({k})}(\varphi^{(n)},\psi^{(n)})
W^n({\Vcy} |{\Vcx} (k))
\notag\\
& \qquad +\frac{1}{|{\cal K}_n|}
\sum_{k \in {\cal K}_n}
\sum_{V \in {\cal V}_n({\cal Y}|P)}
\sum_{\scs {\lvc y}  \in 
\left[\bigcup_{m \neq k}{\cal T}_{\gamma,2,V}^{(n)}({\Vcx} (m))\right]^{\rm c}
\atop{\scs 
\bigcap {\cal T}_{\gamma,2,V}^{(n)}({\Vcx} (k))
      }
}1
\notag\\
& \qquad \times 
\chi_{{\lvc y} |{\Vcx} ({k})}(\varphi^{(n)},\psi^{(n)})
W^n({\Vcy} |{\Vcx} (k))
\notag\\
&\MLeq{a}
\frac{1}{|{\cal K}_n|}
\sum_{k \in {\cal K}_n}
\sum_{V \in {\cal V}_n({\cal Y}|P)}
\sum_{\scs {\lvc y}  \in 
\left[\bigcup_{m \neq k}{\cal T}_{\gamma,2,V}^{(n)}({\Vcx} (m))\right]
\atop{\scs 
\bigcap {\cal T}_{\gamma,2,V}^{(n)}({\Vcx} (k))
      }
}1
\notag\\
& \qquad \times 
W^n({\Vcy} |{\Vcx} (k))
\notag\\
&\leq 
\frac{1}{|{\cal K}_n|}
\sum_{k \in {\cal K}_n}
\sum_{V \in {\cal V}_n({\cal Y}|P)}
\sum_{m \neq k}
\sum_{\scs {\lvc y}  \in {\cal T}_{\gamma,2,V}^{(n)}({\Vcx} (m))}1
\notag\\
& \qquad \times W^n({\Vcy} |{\Vcx} (k)).
\label{eqn:ZssWW}
\end{align}
Step (a) follows from that if
$$
{\Vcy}  \in 
\left[\bigcup_{m \neq k}{\cal T}_{\gamma,2,V}^{(n)}({\Vcx} (m))\right]^{\rm c}
\bigcap {\cal T}_{\gamma,2,V}^{(n)}({\Vcx} (k)),
$$
then the decoding errors do not occur. Set 
\beqno
\zeta_{V}({\Vcx} (k))&\defeq&
\sum_{m \neq k}
\sum_{\scs {\lvc y}  \in {\cal T}_{\gamma,2,V}^{(n)}({\Vcx} (m))}
W^n({\Vcy} |{\Vcx} (k)).
\eeqno
Then from (\ref{eqn:ZssWW}), we have
\beqa
\Lambda_2& \leq &\frac{1}{|{\cal K}_n|}
\sum_{k \in {\cal K}_n}
\sum_{V \in {\cal V}_n({\cal Y}|P)}
\zeta_{V}({\Vcx} (k)).
\label{eqn:SddXXx}
\eeqa
Taking expactations of both sides of (\ref{eqn:SddXXx}) with respect 
to the randomeness of the choice of $\varphi^{(n)}$, we obtain
\beqa
{\sf E}[\Lambda_2]& \leq &\frac{1}{|{\cal K}_n|}
\sum_{k \in {\cal K}_n}
\sum_{V \in {\cal V}_n({\cal Y}|P)}
{\sf E}[\zeta_{V}({\Vcx} (k))].
\label{eqn:SddXXxsss}
\eeqa
For each $ k\in {\cal K}_n$, we evaluate ${\sf E}[\zeta_{V}({\Vcx} (k))]$
to obtain the following chain of inequalities:
\begin{align}
& {\sf E}[\zeta_{V}({\Vcx} (k))]
=\sum_{m \neq k}
\sum_{{\lvc x}(m) \in T^n_P} \sum_{{\lvc x}(k) \in T^n_P} 
\sum_{\scs {\lvc y}  \in {\cal T}_{\gamma,2,V}^{(n)}({\lvc x} (m))}1
\notag\\
&\quad \times W^n({\Vcy} |{\Vcx} (k))
{\sf P}(\varphi^{(n)}(m)={\vc x}(m),\varphi^{(n)}(k)={\vc x}(k))
\notag\\
&=\sum_{m \neq k}
\sum_{{\lvc x} (m) \in T^n_P}\frac{1}{|T^n_P|}
\sum_{\scs {\lvc y}  \in {\cal T}_{\gamma,2,V}^{(n)}({\lvc x} (m))}1
\notag\\
&\quad \times \sum_{{\lvc x}(k) \in T^n_P} 
W^n({\Vcy} |{\Vcx} (k))\frac{1}{|T^n_P|}
\notag\\
&\MEq{a}\sum_{m \neq k}
\sum_{\scs {\lvc x} (m) \in T^n_P}\frac{1}{|T^n_P|}
\sum_{\scs {\lvc y}  \in {\cal T}_{\gamma,2,V}^{(n)}({\lvc x} (m))}Q({\Vcy} )
\notag\\
&\MLeq{b}
\sum_{m \neq k}
\sum_{{\lvc x}(m) \in T^n_P}\frac{1}{|T^n_P|}
\sum_{\scs {\lvc y}  \in {\cal T}_{\gamma,2,V}^{(n)}({\lvc x} (m))}1
\notag\\
&\quad \times 
\kappa_n(|{\cal X}|) 2^{-n J ({\lvc x} (m);{\lvc y}|W)}W^n({\Vcy} |{\Vcx} (m))
\notag\\
&\MLeq{c}
\sum_{m \neq k}
\sum_{{\lvc x} (m) \in T^n_P}\frac{1}{|T^n_P|}
\sum_{\scs {\lvc y}  \in {\cal T}_{\gamma,2,V}^{(n)}({\lvc x} (m))}1
\notag\\
&\quad \times 
\frac{\kappa_n(|{\cal X}|)}
{|{\cal K}_n|2^{n\gamma}}
W^n({\Vcy} |{\Vcx} (m))
\notag\\
&=\frac{\kappa_n(|{\cal X}|)}{|{\cal K}_n|2^{n\gamma}}
\sum_{m \neq k}
\sum_{{\lvc x} (m) \in T^n_P}\frac{1}{|T^n_P|}
W^n({\cal T}_{\gamma,2,V}^{(n)}({\Vcx} (m))|{\Vcx} (m))
\notag\\
&\leq \frac{\kappa_n(|{\cal X}|)}{|{\cal K}_n|2^{n\gamma}}
\sum_{m \neq k}
\sum_{{\lvc x} (m) \in T^n_P}\frac{1}{|T^n_P|}
\notag\\
&=\frac{ \kappa_n(|{\cal X}|)(|{\cal K}_n|-1)}{|{\cal K}_n|2^{n\gamma}}
\leq   \kappa_n(|{\cal X}|) 2^{-n\gamma}.
\label{eqn:0SddXX}
\end{align}
Step (a) follows from the definition of $Q({\vc y})$.
Step (b) follows from Lemma \ref{lm:LemmaA}.
Step (c) follows from that when 
${\vc y}\in {\cal T}_{\gamma,2}({\Vcx}(m))$, we have 
$$
2^{-n J({\lvc x} (m);{\lvc y}|W)}\leq 
\frac{1}{|{\cal K}_n|2^{n \gamma }}.
$$
From (\ref{eqn:SddXXxsss}) and (\ref{eqn:0SddXX}), we have 
${\sf E}[\Lambda_2]\leq \kappa_n(|{\cal X}|) {\ExP}^{-n \gamma}$. 
Hence there exists at least one deterministic code 
such that $\Lambda_2\leq \kappa_n(|{\cal X}|) {\ExP}^{-n \gamma}$. 
Thus we have 
\begin{align*}
&{\rm P}_{\rm e}^{(n)}(\varphi^{(n)},\psi^{(n)}|W)
\notag\\
&\leq 
{\rm Pr}\left\{
 \frac{1}{n}\log |{\cal K}_n| \geq 
J(P,V_{Y^n|\varphi^{(n)}(K_n)}|W)-\gamma\right\}
\notag \\
&\qquad +\kappa_n(|{\cal X}|){\ExP}^{-n \gamma}, 
\end{align*}
completing the proof.
\hfill\IEEEQED 

\newcommand{\SDeee}{
}{

We finally prove Proposition \ref{pro:pro3}. 
For received sequence ${\Vcy} \in {\cal Y}^n$, we 
define the decoder function by
$$
\psi^{(n)}({\Vcy} )
=\left\{
\ba{cl}
\hat{k} 
&\mbox{ if }I({\Vcx} (k) 
;{\Vcy} )
<I({\Vcx} (\tilde{k})
;{\Vcy} )
\\
&\mbox{ for all } \tilde{k}\in {\cal K}_n-\{\hat{k}\},
\\
\mbox{0}&\mbox{ otherwise.}
\ea
\right.
$$

\noindent
\underline{Error Probability Analysis:} 
For ${\Vcy}  \in {\cal Y}^n$, we set
\beqno
{\cal F}({\Vcx} |{\Vcy} )
&\defeq&
\left\{\tilde{\Vcx} \in {\cal X}^n:
I(\tilde{\Vcx};{\Vcy} )\geq I({\Vcx} ;{\Vcy} )\right\}
\\
&=&
\left\{\tilde{\Vcx} \in {\cal X}^n:
H(\tilde{\Vcx}|{\Vcy} )\leq H({\Vcx} |{\Vcy} )\right\}.
\eeqno

Let ${\sf P}$ be a probability measure based on the 
randomness of the choice of $\{{\Vcx} (k)\}_{k\in {\cal K}_n}$. 
For each $k\in {\cal K}_n$ and ${\Vcy}  \in {\cal Y}^n$, 
we consider the following event.
$$
\ba{ll}
{\cal E}_{k}({\Vcy} ): &\:\mbox{There exists }
\tilde{k} \in {\cal K}_n-\{k\}\mbox{ such that}
\\
&\:{\Vcx} (\tilde{k})\in {\cal F}({\Vcx} (k)|{\Vcy} ).
\ea
$$
Then we have the following lemma.
\begin{lm}\label{lm:Sddxxx}
For each $k\in {\cal K}_n$ and ${\Vcy}  \in {\cal Y}^n$, we have
\beqno
{\sf P}({\cal E}_{k}({\Vcy} ))
& &\leq 
\eta_n(|{\cal X}|,|{\cal Y}|)
{\ExP}^{-n 
\left[ I(P,V_{{\svc y} |{\svc x}(k)})-(1/n)\log |{\cal K}_n| \right]},
\eeqno
where 
\end{lm}

{\it Proof:} We first bound the cardinality of ${\cal F}({\Vcx} (k)|{\Vcy} )$. 
On this quantity we have the following:
\beqa
& &|{\cal F}({\Vcx} (k)|{\Vcy} )|
\nonumber\\
&=&|\left\{\tilde{\Vcx} \in {\cal X}^n:
H(\tilde{\Vcx}|{\Vcy} )\leq H({\Vcx} |{\Vcy} )\right\}|
\nonumber\\
&=&\sum_{\scs V \in {\cal V}({\cal X}|P_{{\svc y}}):
\atop{\scs H(V|P_{{\svc y}})\quad
    \atop {\scs \leq H({\lvc x}(k)|{\lvc y})
      }
   }
}
|T_V^n({\Vcy})|
\MLeq{a} 
\sum_{\scs V \in {\cal V}({\cal X}|P_{{\svc y} }):
\atop{\scs H(V|P_{{\svc y} })
    \atop {\scs \leq H({\lvc x}(k)|{\lvc y})
      }
   }
}
{\ExP}^{nH(V|P_{{\svc y} })}
\nonumber\\
&\leq& {\ExP}^{nH({\lvc x} (k)|{\lvc y} )}
\sum_{\scs V \in {\cal V}({\cal X}|P_{{\svc y} }):
\atop{\scs H(V|P_{{\svc y} })
    \atop {\scs \leq H({\lvc x}(k)|{\lvc y})
      }
   }
}1
\nonumber\\
&\leq & \eta_n(|{\cal X}|,|{\cal Y}|)
{\ExP}^{nH({\lvc x}(k)|{\lvc y} )}.
\label{eqn:aSgg}
\eeqa
Step (a) follows from Lemma \ref{lm:Lem1} part a). 
Step (b) follows from Lemma \ref{lm:Lem1} part b). 
Then we have 
\beqno
& &{\sf P}({\cal E}_{k}({\Vcy} ))
\leq (|{\cal K}_n|-1) 
\sum_{\tilde{\lvc x} \in {\cal F}({\lvc x}(k)|{\lvc y} )}
\frac{1}{|T^n_{P}|}
\nonumber\\
&\leq &|{\cal K}_n|\frac{ |{\cal F}({\Vcx} (k)|{\Vcy})|}{|T^n_{P_{{\svc x} }}|}
\MLeq{a}
|{\cal K}_n|
\frac{\nu_n(|{\cal X}||{\cal Y}|)
\ExP^{nH( {\lvc x}(k)|{{\lvc y} })}}{[\kappa_n(|{\cal X}|)]^{-1}
\ExP^{nH({{\lvc x} })}}
\nonumber\\
&=&
\eta_n(|{\cal X}|,|{\cal Y}|)\ExP^{-n\left[I(P,V_{{\svc y} |{\svc x}(k)})-(1/n)
\log |{\cal K}_n|\right]}.
\eeqno
Step (a) follows from Lemma \ref{lm:Lem1} and (\ref{eqn:aSgg}).
\hfill \IEEEQED
}

{\it Proof of Proposition \ref{pro:pro3}:} We take 
our proposed encoder and decoder functions as 
$(\varphi^{(n)},\psi^{(n)})$. 
By the construction of $\varphi^{(n)}$, we have that
for any $k\in {\cal K}_n$, ${\Vcx} (k) \in T^n_P.$
For $k\in {\cal K}_n$ and 
${\cal V}_n({\cal Y}|P)$, set
\beqno
{\cal T}_{k,V}^{(1)}&\defeq& \left\{ {\Vcy}  \in T_V^n({\Vcx} (k)):
\frac 1 n \log |{\cal K}_n| \geq I(P,V) -\gamma \right\},
\\
{\cal T}_{k,V}^{(2)}&\defeq& \left\{ {\Vcy}  \in T_V^n({\Vcx} (k)):
\frac 1 n \log |{\cal K}_n| < I(P,V) -\gamma \right\}.
\eeqno
For $k\in {\cal K}_n$ and 
for $({\Vcx} (k),{\Vcy} )\in {\cal X}^n\times {\cal Y}^n$, 
define
$$
\chi_{{\lvc y} |{\lvc x} (k)}(\varphi^{(n)},\psi^{(n)})
\defeq\left\{ 
\ba{ll} 1 &\mbox{ if }\psi^{(n)}({\Vcy} )\neq k,
\\
 0 & \mbox{ otherwise. }
\ea
\right.
$$
Then we have 
${\rm P}_{\rm e}^{(n)}(\varphi^{(n)},\psi^{(n)}|W)
\newcommand{\omitA}{
&=&\frac{1}{|{\cal K}_n|}
\sum_{k\in {\cal K}_n}
\sum_{{\lvc y} \in {\cal Y}^n}
\chi_{{\lvc y} |{\lvc x} ({k})}(\varphi^{(n)},\psi^{(n)})
W^n({\Vcy} |{\Vcx} (k))
\\
&=&
\frac{1}{|{\cal K}_n|}
\sum_{k\in {\cal K}_n}
\sum_{V \in {\cal V}_n({\cal Y}|P_{{\lvc x} (k)})}
\sum_{{\lvc y}  \in T_V^n({\lvc x} )}1
\\
& &\qquad \times \chi_{{\lvc y} |{\lvc x} ({k})}(\varphi^{(n)},\psi^{(n)})
W^n({\Vcy} |{\Vcx} (k))
\\
&=&\Lambda_1+\Lambda_2, 
}
=\Lambda_1+\Lambda_2, 
$
where
\beqno
\Lambda_1&=&\frac{1}{|{\cal K}_n|}
\sum_{k \in {\cal K}_n}
\sum_{V \in {\cal V}_n({\cal Y}|P)}
\sum_{{\lvc y}  \in {\cal T}_{k,V}^{(1)}({\lvc x} (k))}1
\\
& &\qquad\times 
\chi_{{\lvc y} |{\lvc x} ({k})}(\varphi^{(n)},\psi^{(n)})W^n({\Vcy} |{\Vcx} (k)),
\\
\Lambda_2&=&\frac{1}{|{\cal K}_n|}
\sum_{k \in {\cal K}_n}
\sum_{V \in {\cal V}_n({\cal Y}|P)}
\sum_{{\lvc y}  \in {\cal T}_{k,V}^{(2)}({\lvc x} (k))}1
\\
& &\qquad \times 
\chi_{{\lvc y} |{\lvc x} ({k})}(\varphi^{(n)},\psi^{(n)})W^n({\Vcy} |{\Vcx} (k)).
\eeqno
On an upper bound of $\Lambda_1$, we have the following:
\begin{align*}
&    \Lambda_1
 \leq  \frac{1}{|{\cal K}_n|}
\sum_{k \in {\cal K}_n}
\sum_{V \in {\cal V}_n({\cal Y}|P)}
\sum_{{\lvc y}  \in {\cal T}_{k,V}^{(1)}({\lvc x} (k))}
W^n({\Vcy} |{\Vcx} (k))
\\
&=
{\rm Pr}\left\{
 \frac{1}{n}\log |{\cal K}_n| \geq 
I(P,V_{Y^n|\varphi^{(n)}(K_n)})-\gamma\right\}.
\end{align*} 
We next derive an upper bound of $\Lambda_2$. 
Let ${\sf E}$ denote an expectation based on a randomness 
of the choice of $\varphi^{(n)}$. 
\newcommand{\OmitZ}{}
{
We evaluate 
${\sf E}[\Lambda_2]$ to obtain the following chain of 
inequalities: 
\begin{align*}
& {\sf E}[\Lambda_2]
={\sf E}\Hugebl \frac{1}{|{\cal K}_n|}
\sum_{k \in {\cal K}_n}
\sum_{V \in {\cal V}_n({\cal Y}|P)}
\sum_{{\lvc y}  \in {\cal T}_{k,V}^{(2)}({\lvc x} (k))}1
\\
& \qquad\times 
\chi_{{\lvc y} |{\lvc x} ({k})}(\varphi^{(n)},\psi^{(n)})
W^n({\Vcy} |{\Vcx} (k)) \Hugebr
\\
&=\frac{1}{|{\cal K}_n|}
\sum_{k \in {\cal K}_n}
\sum_{V \in {\cal V}_n({\cal Y}|P)}
\sum_{{\lvc y}  \in {\cal T}_{k,V}^{(2)}({\lvc x} (k))}1
\\
& \qquad\times 
{\sf E}\left[\chi_{{\lvc y} |{\lvc x} ({k})}
(\varphi^{(n)},\psi^{(n)})\right]W^n({\Vcy} |{\Vcx} (k))
\\
&=\frac{1}{|{\cal K}_n|}
\sum_{k \in {\cal K}_n}
\sum_{V \in {\cal V}_n({\cal Y}|P)}
\sum_{{\lvc y}  \in {\cal T}_{k,V}^{(2)}({\lvc x} (k))}1
\\
& \qquad\times 
{\sf P}({\cal E}_{k}({\Vcy} ))W^n({\Vcy} |{\Vcx} (k))
\\
&\MLeq{a}
\frac{1}{|{\cal K}_n|}
\sum_{k \in {\cal K}_n}
\sum_{V \in {\cal V}_n({\cal Y}|P)}
\sum_{{\lvc y}  \in {\cal T}_{k,V}^{(2)}({\lvc x} (k))}W^n({\Vcy} |{\Vcx} (k))
\\
& \qquad \times 
\eta_n(|{\cal X}|,|{\cal Y}|)
{\ExP}^{-n \left[ I(P,V_{{\svc y} |{\svc x}(k)})-(1/n)\log |{\cal K}_n| \right]}
\\
&\MLeq{b}
\frac{1}{|{\cal K}_n|}
\sum_{k \in {\cal K}_n}
\sum_{V \in {\cal V}_n({\cal Y}|P)}
\sum_{{\lvc y}  \in {\cal T}_{k,V}^{(2)}({\lvc x} (k))}W^n({\Vcy} |{\Vcx} (k))
\\
& \qquad \times \eta_n(|{\cal X}|,|{\cal Y}|){\ExP}^{-n \gamma}
\\
& \leq  \eta_n(|{\cal X}|,|{\cal Y}|) {\ExP}^{-n \gamma}.
\end{align*}
Step (a) follows from Lemma \ref{lm:Sddxxx}. 
Step (b) follows from that when 
${\Vcy}  \in $ ${\cal T}_{k,V}^{(2)}({\Vcx} (k))$, 
we have 
$$
{\ExP}^{-n \left[ I(P,V_{{\svc y} |{\svc x}(k)})-(1/n)\log |{\cal K}_n| \right]}
\leq {\ExP}^{-n \gamma}.
$$
}
Hence there exists at least one deterministic code 
such that $\Lambda_2\leq \eta_n{\ExP}^{-n \gamma}$. Thus we have 
\begin{align*}
&{\rm P}_{\rm e}^{(n)}(\varphi^{(n)},\psi^{(n)}|W)
\notag\\
&\leq 
{\rm Pr}\left\{
 \frac{1}{n}\log |{\cal K}_n| \geq 
I(P,V_{Y^n|\varphi^{(n)}(K_n)})-\gamma\right\}
\notag \\
&\qquad 
+\eta_n(|{\cal X}|,|{\cal Y}|){\ExP}^{-n \gamma}, 
\end{align*}
completing the proof.
\hfill\IEEEQED 

\ProofThs

\appendix

\noindent

\ApdaAAAA

\ReF

\begin{thebibliography}{99}

\bibitem{Strassen62}V. Strassen, ``Asymptotische 
absch\"atungen in Shannon's 
informationstheorie," in {\it Proc. Trans. 3rd 
Prague Conf. Inf. Theory,} Prague, Czech Republic, 
1962, pp. 689-723.  

\bibitem{Hayashi10}M. Hayashi, ``Information spectrum 
approach to second-order coding rate in channel coding," 
IEEE Trans. Inform. Theory, vol. 55, no. 11, pp. 4947--4966, 
Nov. 2009.

\bibitem{PolyanskiyEtal10}
Y. Polyanskiy, H. V. Poor, and S. Verdu, ``Channel coding rate 
in the finite blocklength regime," {\it IEEE Trans. Inf. Theory}, 
vol. 56, no. 5, pp. 2307--2358, May 2010. 

\bibitem{IngberKoch11}
A. Ingber and Y. Kochman, ``The dispersion of lossy source coding,h 
{\it in Proc. Data Compression Conference}, pp. 53--62, 2011.

\bibitem{WangEtal11}
D. Wang, A. Ingber, and Y. Kochman, ``The dispersion of 
joint source channel coding," {\it in Proc. Allerton Conference on 
Communication, Control and Computing}, pp. 180--187, 2011.

\bibitem{Kostina12}
V. Kostina and S. Verdu, ``Fixed-length lossy compression in the finite
blocklength regime," {\it IEEE Trans. Inf. Theory}, 
vol. 58, no. 6, pp. 3309--3338, Jun. 2012.

\bibitem{Kostina13}
V. Kostina and S. Verdu, ``Lossy joint source-channel coding 
in the finite blocklength regime," {\it IEEE Trans. Inf. Theory}, 
vol. 59, no. 5, pp. 2545--2575, May 2013.

\bibitem{Han98InfSpec}
T. S. Han, {\it Information-Spectrum Methods in Information
Theory.} Springer-Verlag, Berlin, New York, 2002. The Japanese 
edition was published by Baifukan-publisher, Tokyo, 1998.

\bibitem{ck} 
I. Csisz\'ar and J. K\"orner, {\it Information Theory: Coding 
Theorems for Discrete Memoryless Systems.} Academic Press, 
New York, 1981.
\end{thebibliography}

\begin{thebibliography}{99}

\bibitem{cov72}T.~M. Cover, ``Broadcast channels,'' 
{\em IEEE Trans. Inform. Theory}, vol. IT-18, 
no.1, pp.~2--13, Jan. 1972.

\bibitem{bgm73}P. P. Bergmans, ``Random coding theorems 
for broadcast channels with degraded components," 
{\it IEEE Trans. Inform. Theory}, vol. IT-19, pp. 197-207, 
Mar. 1973.


\bibitem{gal74}
R. G. Gallager, ``Capacity and coding for degraded broadcast channels," 
{\it Problemy Peredachi Informatsii}, vol. 10, pp. 3-14, 
July-Sept. 1974.

\bibitem{ak75}
R. F. Ahlswede and J. K\"orner, ``Source coding with side information
and a converse for degraded broadcast channels," 
{\it IEEE Trans. Inform. Theory}, vol. IT-21, pp. 629-637, Nov. 1975.



\bibitem{agk76}R. Ahlswede, P. G\`as, and J. K\"orner, 
``Bounds on conditional probabilities with applications 
in multi-user communication," {\it Z. Wahrscheinlichkeitstheorie 
verw. Gebiete}, vol. 34, pp. 157-177, 1976.






\bibitem{ck} I. Csisz\'ar and J. K\"orner, 
{\it Information Theory: Coding Theorems for Discrete 
Memoryless Systems.} Academic Press, New York, 1981.




\bibitem{Han98InfSpec}
T. S. Han, {\it Information-Spectrum Methods in Information
Theory. }Springer-Verlag, Berlin, New York, 2002. The Japanese 
edition was published by Baifukan-publisher, Tokyo, 1998.

\bibitem{ari}S. Arimoto,
``On the converse to the coding theorem for discrete memoryless 
channels,'' {\it IEEE Trans. Inform. Theory,} vol. IT-19, no. 3, pp. 
357-359, May 1973.

\bibitem{Dueck_Korner1979}
G.~Dueck and J.~K\"orner, ``Reliability function of a discrete 
memoryless channel at rates above capacity,'' 
\emph{IEEE Trans. Inform. Theory}, vol. IT-25, no.~1, pp. 
82--85, 1979.


\bibitem{ElGamalBcFb78}
A. B. El Gamal, ``The feedback capacity of degraded broadcast channels,"
{\it IEEE Trans. Inform. Theory}, vol. IT-24, no.3, pp.379-381, May 1978.

\bibitem{ShaWi13BcFb13} O. Shayevitz and M. Wigger, 
``On the Capacity of the discrete memoryless broadcast 
channel with feedback, {\it IEEE Trans. Inform. Theory}, 
vol. 59, no. 3, March 2013, 1329-1345.

\bibitem{ckf}I. Csisz\'ar and J. K\"orner, 
``Feedback does not affect the reliability function of 
a DMC at rates above capacity,''
{\it IEEE Trans. Inform. Theory}, vol.IT-28, pp.92-93, 1982.

\end{thebibliography}
\end{document}